
\def\singlespace{\normalbaselines}
\def\oneandahalfspace{\baselineskip=1.15\normalbaselineskip plus 1pt
\lineskip=2pt\lineskiplimit=1pt}

\def\np{\vfill\eject}
\def\nl{\hfil\break}

\def\nofirstpagenoten{\nopagenumbers\footline={\ifnum\pageno>1\tenrm
\hss\folio\hss\fi}}
\def\nofirstpagenotwelve{\nopagenumbers\footline={\ifnum\pageno>1\twelverm
\hss\folio\hss\fi}}
\def\leaderfill{\leaders\hbox to 1em{\hss.\hss}\hfill}
\def\ft#1#2{{\textstyle{{#1}\over{#2}}}}
\def\frac#1/#2{\leavevmode\kern.1em
\raise.5ex\hbox{\the\scriptfont0 #1}\kern-.1em/\kern-.15em
\lower.25ex\hbox{\the\scriptfont0 #2}}
\def\sfrac#1/#2{\leavevmode\kern.1em
\raise.5ex\hbox{\the\scriptscriptfont0 #1}\kern-.1em/\kern-.15em
\lower.25ex\hbox{\the\scriptscriptfont0 #2}}


\parindent=20pt
\def\narrow{\advance\leftskip by 40pt \advance\rightskip by 40pt}

\def\AB{\bigskip
        \centerline{\bf ABSTRACT}\medskip\narrow}
\def\nonarrower{\advance\leftskip by -40pt\advance\rightskip by -40pt}
\def\AE{\bigskip\nonarrower}

\def\boxit#1{\vbox{\hrule\hbox{\vrule\kern3pt
        \vbox{\kern3pt#1\kern3pt}\kern3pt\vrule}\hrule}}

\def\gtorder{\mathrel{\raise.3ex\hbox{$>$}\mkern-14mu
             \lower0.6ex\hbox{$\sim$}}}
\def\ltorder{\mathrel{\raise.3ex\hbox{$<$}|mkern-14mu
             \lower0.6ex\hbox{\sim$}}}
\def\dalemb#1#2{{\vbox{\hrule height .#2pt
        \hbox{\vrule width.#2pt height#1pt \kern#1pt
                \vrule width.#2pt}
        \hrule height.#2pt}}}

\font\fourteentt=cmtt10 scaled \magstep2
\font\fourteenbf=cmbx12 scaled \magstep1
\font\fourteenrm=cmr12 scaled \magstep1
\font\fourteeni=cmmi12 scaled \magstep1
\font\fourteenss=cmss12 scaled \magstep1
\font\fourteensy=cmsy10 scaled \magstep2
\font\fourteensl=cmsl12 scaled \magstep1
\font\fourteenex=cmex10 scaled \magstep2
\font\fourteenit=cmti12 scaled \magstep1
\font\twelvett=cmtt10 scaled \magstep1 \font\twelvebf=cmbx12
\font\twelverm=cmr12 \font\twelvei=cmmi12
\font\twelvess=cmss12 \font\twelvesy=cmsy10 scaled \magstep1
\font\twelvesl=cmsl12 \font\twelveex=cmex10 scaled \magstep1
\font\twelveit=cmti12
\font\tenss=cmss10
 
 \font\ninebf=cmbx7 scaled \magstep1
\font\ninerm=cmr7 scaled \magstep1 \font\ninei=cmmi7 scaled \magstep1
\font\ninesy=cmsy7 scaled \magstep1 
\font\eightrm=cmr7 scaled 1140 
 
\font\sevenbf=cmbx7 \font\sevenrm=cmr7 \font\seveni=cmmi7
\font\sevensy=cmsy7 

\catcode`@=11
\newskip\ttglue
\newfam\ssfam

\def\fourteenpoint{\def\rm{\fam0\fourteenrm}
\textfont0=\fourteenrm \scriptfont0=\tenrm \scriptscriptfont0=\sevenrm
\textfont1=\fourteeni \scriptfont1=\teni \scriptscriptfont1=\seveni
\textfont2=\fourteensy \scriptfont2=\tensy \scriptscriptfont2=\sevensy
\textfont3=\fourteenex \scriptfont3=\fourteenex \scriptscriptfont3=\fourteenex
\def\it{\fam\itfam\fourteenit} \textfont\itfam=\fourteenit
\def\sl{\fam\slfam\fourteensl} \textfont\slfam=\fourteensl
\def\bf{\fam\bffam\fourteenbf} \textfont\bffam=\fourteenbf
\scriptfont\bffam=\tenbf \scriptscriptfont\bffam=\sevenbf
\def\tt{\fam\ttfam\fourteentt} \textfont\ttfam=\fourteentt
\def\ss{\fam\ssfam\fourteenss} \textfont\ssfam=\fourteenss
\tt \ttglue=.5em plus .25em minus .15em
\normalbaselineskip=16pt
\abovedisplayskip=16pt plus 4pt minus 12pt
\belowdisplayskip=16pt plus 4pt minus 12pt
\abovedisplayshortskip=0pt plus 4pt
\belowdisplayshortskip=9pt plus 4pt minus 6pt
\parskip=5pt plus 1.5pt
\setbox\strutbox=\hbox{\vrule height12pt depth5pt width0pt}
\let\sc=\tenrm
\let\big=\fourteenbig \normalbaselines\rm}
\def\fourteenbig#1{{\hbox{$\left#1\vbox to12pt{}\right.\n@space$}}}

\def\twelvepoint{\def\rm{\fam0\twelverm}
\textfont0=\twelverm \scriptfont0=\ninerm \scriptscriptfont0=\sevenrm
\textfont1=\twelvei \scriptfont1=\ninei \scriptscriptfont1=\seveni
\textfont2=\twelvesy \scriptfont2=\ninesy \scriptscriptfont2=\sevensy
\textfont3=\twelveex \scriptfont3=\twelveex \scriptscriptfont3=\twelveex
\def\it{\fam\itfam\twelveit} \textfont\itfam=\twelveit
\def\sl{\fam\slfam\twelvesl} \textfont\slfam=\twelvesl
\def\bf{\fam\bffam\twelvebf} \textfont\bffam=\twelvebf
\scriptfont\bffam=\ninebf \scriptscriptfont\bffam=\sevenbf
\def\tt{\fam\ttfam\twelvett} \textfont\ttfam=\twelvett
\def\ss{\fam\ssfam\twelvess} \textfont\ssfam=\twelvess
\tt \ttglue=.5em plus .25em minus .15em
\normalbaselineskip=14pt
\abovedisplayskip=14pt plus 3pt minus 10pt
\belowdisplayskip=14pt plus 3pt minus 10pt
\abovedisplayshortskip=0pt plus 3pt
\belowdisplayshortskip=8pt plus 3pt minus 5pt
\parskip=3pt plus 1.5pt
\setbox\strutbox=\hbox{\vrule height10pt depth4pt width0pt}
\let\sc=\ninerm
\let\big=\twelvebig \normalbaselines\rm}
\def\twelvebig#1{{\hbox{$\left#1\vbox to10pt{}\right.\n@space$}}}

\def\tenpoint{\def\rm{\fam0\tenrm}
\textfont0=\tenrm \scriptfont0=\sevenrm \scriptscriptfont0=\fiverm
\textfont1=\teni \scriptfont1=\seveni \scriptscriptfont1=\fivei
\textfont2=\tensy \scriptfont2=\sevensy \scriptscriptfont2=\fivesy
\textfont3=\tenex \scriptfont3=\tenex \scriptscriptfont3=\tenex
\def\it{\fam\itfam\tenit} \textfont\itfam=\tenit
\def\sl{\fam\slfam\tensl} \textfont\slfam=\tensl
\def\bf{\fam\bffam\tenbf} \textfont\bffam=\tenbf
\scriptfont\bffam=\sevenbf \scriptscriptfont\bffam=\fivebf
\def\tt{\fam\ttfam\tentt} \textfont\ttfam=\tentt
\def\ss{\fam\ssfam\tenss} \textfont\ssfam=\tenss
\tt \ttglue=.5em plus .25em minus .15em
\normalbaselineskip=12pt
\abovedisplayskip=12pt plus 3pt minus 9pt
\belowdisplayskip=12pt plus 3pt minus 9pt
\abovedisplayshortskip=0pt plus 3pt
\belowdisplayshortskip=7pt plus 3pt minus 4pt
\parskip=0.0pt plus 1.0pt
\setbox\strutbox=\hbox{\vrule height8.5pt depth3.5pt width0pt}
\let\sc=\eightrm
\let\big=\tenbig \normalbaselines\rm}
\def\tenbig#1{{\hbox{$\left#1\vbox to8.5pt{}\right.\n@space$}}}
\let\rawfootnote=\footnote \def\footnote#1#2{{\rm\parskip=0pt\rawfootnote{#1}
{#2\hfill\vrule height 0pt depth 6pt width 0pt}}}

\def\tenfoot{\tenpoint\hskip-\parindent\hskip-.1cm}

\overfullrule=0pt
\twelvepoint
\def\sbullet{\raise.2em\hbox{$\scriptscriptstyle\bullet$}}
\nofirstpagenotwelve
\hsize=16.5 truecm
\baselineskip 15pt

\def\ft#1#2{{\textstyle{{#1}\over{#2}}}}

\def\ket#1{\big| #1\big\rangle}
\def\bra#1{\big\langle #1\big|}

\def\dum{{\phantom{X}}}
\def\m{{\rm -}}

\def\phys{\big|{\rm phys}\big\rangle}
\def\fpf#1#2#3#4{\big\langle #1\ #2\ #3\ #4\big\rangle}
\def\fipf#1#2#3#4#5{\big\langle #1\ #2\ #3\ #4\ #5\big\rangle}
\def\spf#1#2#3#4#5#6{\big\langle #1\ #2\ #3\ #4\ #5\ #6\big\rangle}
\def\deff{\Delta}
\def\teff{T^{\rm eff}}
\def\leff{L^{\rm eff}}
\def\V#1#2#3#4{{\bf V}^{#1}_{#2}[#3,#4]}
\def\aV#1#2#3#4{{\bf W}^{#1}_{#2}[#3,#4]}
\def\D#1#2{{\bf D}^{#1}[#2]}

\def\b{\beta}

\def\del{\partial}

\def\phys{\big|\hbox{phys}\big\rangle}

\oneandahalfspace
\rightline{CTP TAMU--4/93}
\rightline{Preprint-KUL-TF-93/2}
\rightline{hep-th/9301099}
\rightline{January 1993}

\vskip 2truecm
\centerline{\bf On the Spectrum and Scattering of $W_3$ Strings}
\vskip 1.5truecm
\centerline{H. Lu, C.N. Pope,\footnote{$^*$}{\tenfoot Supported in part
by the U.S. Department of Energy, under
grant DE-FG05-91ER40633.} S. Schrans\footnote{$^\diamond$}{\tenfoot
Onderzoeker I.I.K.W.;
On leave of absence from the Instituut voor Theoretische Fysica, \nl
\indent$\,$ K.U. Leuven, Belgium. Address after March 1, 1993:
Koninklijke/Shell-Laboratorium, \nl
\indent$\,$ Amsterdam (Shell Research B.V.),
Badhuisweg  3, 1031 CM Amsterdam, The Netherlands.
}
and X.J.
Wang\footnote{}{\tenfoot }}
\vskip 1.5truecm
\centerline{\it Center
for Theoretical Physics,
Texas A\&M University,}
\centerline{\it College Station, TX 77843--4242, USA.}

\vskip 1.5truecm
\AB\singlespace
    We present a detailed investigation of scattering processes in $W_3$
string theory.  We discover further physical states with continuous
momentum, which involve excitations of the ghosts as well as the matter, and
use them to gain a better understanding of the interacting theory.  The
scattering amplitudes display factorisation properties, with states from the
different sectors of the theory being exchanged in the various intermediate
channels.  We find strong evidence for the unitarity of the theory, despite
the unusual ghost structure of some of the physical states.  Finally, we
show that by performing a transformation of the quantum fields that involves
mixing the ghost fields with one of the matter fields, the structure of the
physical states is dramatically simplified.  The new formalism provides a
concise framework within which to study the $W_3$ string.
\AE\oneandahalfspace

\np
\noindent
{\bf 1. Introduction}
\bigskip

      Since the discovery of $W_3$ symmetry [1] in two-dimensional field
theories much work has been carried out on the construction of $W_3$-string
theories [2--12]. Most of these efforts have been concerned with the
understanding of their physical spectra [2--10,12]. Until recently,
attention had been focussed on physical states with standard ghost
structure, which are direct analogues of the physical states in the usual
26-dimensional bosonic string.   It has been recently realised that there
are also physical states with non-standard ghost structure in the $W_3$
string, {\it i.e.}\ states that involve excitations of the ghost fields as
well as the matter fields [13,9--12].  Unlike the two-dimensional string,
where states of this kind occur too [14,15], they exist not only in the
two-scalar $W_3$ string where the momenta are discrete, but also in the {\it
multi}-scalar $W_3$ string where the momenta are in general continuous.  The
occurrence of states involving excitations of the ghost fields in the
physical spectrum is rather unusual in a gauge theory, and normally one
might expect that it would be associated with a violation of unitarity.
However, as was discussed in [11], and will be explored further in this
paper, it seems that these states {\it are} unitary, and furthermore, they
are an essential ingredient of the $W_3$ string.  In fact, as we recently
showed [11], the existence of physical states with non-standard ghost
structure resolves the long-standing puzzle of how to introduce interactions
in the $W_3$ string.

      As in ordinary string theory, interactions have to be introduced by
hand.  The guiding principle that determines acceptable interactions is that
they should respect the symmetries of the theory.  This means in the present
context that they should lead to BRST-invariant scattering amplitudes.  The
stumbling-block to introducing interactions in the $W_3$ string was that
for a long time only the physical states with standard ghost structure were
known, and with these it appears to be impossible to write down any
non-vanishing BRST-invariant amplitude.  The reason for this is that these
physical states, and indeed {\it all} physical states in the $W_3$ string,
have specific values of momentum in a particular ``frozen'' direction,
and the particular values that occur for the states with standard ghost
structure imply that momentum conservation in this direction cannot be
satisfied in interactions.

     In [11], we presented a procedure for building gauge-invariant scattering
amplitudes in the $W_3$ string.  An essential aspect is that
non-vanishing scattering amplitudes necessarily involve external physical
states with non-standard ghost structure.  The problem with momentum
conservation in the frozen direction is avoided because these states have
different values of frozen momentum from those of the states with standard
ghost structure. In this paper we shall review and elaborate on this
procedure, resolve some open problems in [11], and begin the unravelling of
the general structure of $W_3$-string scattering.

     The paper is organised as follows.  Section 2 contains a summary of the
main results in [11] and outlines several issues that we shall explore further
in this paper.

     We begin section 3 by finding new physical states of the
$W_3$ string; they have higher levels, and lower ghost numbers, than the ones
that were known previously.  All the physical states we know fall
into two categories: those with continuous on-shell spacetime momentum, and
those where the on-shell spacetime momentum can only take a discrete value
(in fact, zero). The new physical states  enable us to gain a better
understanding of the interactions of the $W_3$ string. In particular, we are
now able to construct certain interactions that could not have been obtained
with the physical states of [11] alone.  We then explain how the physical
states can be organised into multiplets and discuss their structure in
detail.   One of the main features that emerges from the examples of
physical states with continuous momentum  that we have constructed is that
they all admit an effective spacetime interpretation. By this we mean that a
physical state of this kind factorises into a direct product of a piece
involving the ghost fields together with the frozen coordinate, and another
piece involving the remaining coordinates which can then be viewed as the
coordinates of the target spacetime.  Moreover, we find that from the
effective spacetime point of view, these states fall into three sectors,
with effective conformal dimension $\leff_0=1$, $\ft{15}{16}$ and $\ft12$.

     In section 4 we use the known physical states to build new
four-point and five-point scattering amplitudes.  One of these is a
four-point function for four physical states in the $\leff_0=1$ sector.  It
provides a useful non-trivial check of our procedure, since it should, and
indeed does, agree with the corresponding result in ordinary bosonic string
theory (at tree level).  We also show that physical states in the
$\leff_0=\ft{15}{16}$ sector play a special r\^ole in the theory in that
there are only very limited possibilities for them to interact with other
states. Specifically, we find that no more than two of the physical states
in a non-vanishing $N$-point function can have $\leff_0=\ft{15}{16}$.

     Section 5 is divided into two subsections.  In the first we make some
general observations about the physical spectrum of the $W_3$ string and
derive necessary conditions for the level numbers and momenta in the frozen
direction of physical states which admit an effective spacetime
interpretation and have $\leff_0$ taking values in the set $\{1,\
\ft{15}{16},\ \ft12\}$.  In the second subsection we consider the unitarity
of the theory.  We show that if all the physical states with continuous
momentum indeed admit an effective spacetime interpretation and have the
above values of $\leff_0$, then the $W_3$ string is unitary.  Further
evidence for unitarity comes from looking at the residues of the poles of
the scattering amplitudes.

    Motivated by the fact that all the continuous-momentum physical states
known so far admit an effective spacetime interpretation, in section 6 we
introduce a formalism that exploits this. It consists of making a certain
transformation of the quantum fields, which mixes the ghost fields
with the frozen coordinate.  It is a canonical transformation, in the sense
that the new fields satisfy the same set of operator-product expansions as
the original ones.   In terms of these redefined fields, the known physical
operators acquire a remarkably simple form. A striking example is provided
by a level 8 physical operator, which reduces from 82 terms to 4 after
transforming to the new fields.  Moreover, this formalism will make more
transparent some of the features of the $W_3$ string that we have
encountered in this paper. The simplification achieved by this
transformation is highly suggestive that the new formalism is the
appropriate one for describing the $W_3$ string.

     The paper ends with concluding remarks in section 7.

\bigskip
\bigskip
\noindent
{\bf 2. A review of the $W_3$ string}
\bigskip

     In this section we review the procedure and main results of [11].  In
order to make the present paper as self-contained as possible, and in order
to establish notation, a certain amount of repetition is unavoidable.

     The key ingredient for determining the physical spectrum of the $W_3$
string is the construction of the BRST operator [16], which is given by
$$
Q_B=\oint dz \Big[c\,(T+\ft12 T_{\rm gh})+\gamma\,(W+\ft12
W_{\rm gh})\Big], \eqno(2.1)
$$
and is nilpotent provided that the matter currents $T$ and $W$ generate the
$W_3$ algebra with central charge $c=100$, and that the ghost currents are
chosen  to be
$$
\eqalignno{
T_{\rm gh}&=-2b\,\partial c-\partial
b\, c-3\beta\, \partial\gamma-2\partial\beta\, \gamma\ , &(2.2)\cr
W_{\rm gh}&=-\partial\beta\,
c-3\beta\, \partial c-\ft8{261}\big[\partial(b\, \gamma\,  T)+b\,
\partial\gamma \, T\big]\cr
&\ \ +\ft{25}{1566}\Big(2\gamma\, \partial^3b+9\partial\gamma\,
\partial^2b +15\partial^2\gamma\,\partial b+10\partial^3\gamma\,
b\Big) ,&(2.3)\cr}
$$
where the ghost-antighost pairs ($c$, $b$) and ($\gamma$, $\beta$) correspond
respectively to the $T$ and $W$ generators. A matter
realisation of $W_3$ with central charge 100 can be given
in terms of $n\ge 2$ scalar fields, as follows [17]:
$$
\eqalign{
T&= -\ft12 (\del\varphi)^2 - Q \del^2 \varphi +\teff,\cr
W&=-{2i \over \sqrt{261} }\Big[ \ft13 (\del\varphi)^3 + Q \del\varphi
\del^2\varphi +\ft13 Q^2 \del^3 \varphi + 2\del\varphi \teff
+ Q \del \teff\Big],\cr}\eqno(2.4)
$$
where $Q^2=\ft{49}{8}$ and $\teff$ is an energy-momentum tensor
with central charge $\ft{51}2$ that commutes with $\varphi$.  Since
$\teff$ has a fractional central charge, it cannot be realised simply by
taking free scalar fields.  We can however use $d$ scalar fields $X^\mu$
with a background-charge vector $a_\mu$:
$$
\teff=-\ft12 \del X_\mu \del X^\mu - i a_\mu \del^2 X^\mu,\eqno(2.5)
$$
with $a_\mu$ chosen so that $\ft{51}2=d - 12 a_\mu a^\mu$ [17].

     Physical states are by definition states that are annihilated by the BRST
operator (2.1) but that are not BRST trivial. Such
states  with standard ghost structure are of the form:
$$
\ket{\chi}=\ket{\m\m} \otimes \ket{\varphi\, ,X}.\eqno(2.6)
$$
Here $\ket{\m\m}$ is the standard ghost vacuum, given by
$$
\ket{\m\m}=c_1\gamma_1\gamma_2\ket{0}.\eqno(2.7)
$$
The $SL(2,C)$ vacuum satisfies
$$
\eqalignno{
c_n\ket{0}=0,\qquad n\ge 2;&\qquad\qquad\qquad
b_n\ket{0}=0,\qquad n\ge -1,&(2.8a)\cr
\gamma_n\ket{0}=0,\qquad n\ge 3;&\qquad\qquad\qquad
\beta_n\ket{0}=0,\qquad n\ge -2.&(2.8b)\cr}
$$
The antighost fields $b$,
$\b$ have ghost number $G=-1$, and the ghost fields $c$, $\gamma$ have
ghost number $G=1$.\footnote{$^*$}{\tenfoot For states, we follow the
convention in [11] that the ghost vacuum $\ket{\m\m}$ has ghost number $G=0$,
which means that the $SL(2,C)$ vacuum $\ket{0}$ has ghost number $G=-3$.
This implies that physical states of ghost number $G$ are obtained by acting
on $\ket{0}$ with operators of ghost number $(G+3)$. }

     For standard states of the form (2.6), the condition of BRST invariance
becomes [16]:
$$
\eqalign{
(L_0-4)\ket{\varphi\, ,X}&=0,\cr
W_0\ket{\varphi\, ,X}&=0,\cr
L_n\ket{\varphi\, ,X}=W_n\ket{\varphi\, ,X}&=0,\qquad n\ge 1.\cr}\eqno(2.9)
$$
The consequences of these physical-state conditions have been studied in
detail in various papers [3--8,10].  The main features that emerge are the
following. There are  two kinds of excited states, namely those
for which there are no excitations in the $\varphi$ direction, and those
where $\varphi$ is excited too. The latter states are all BRST trivial, as
has been discussed in [6,8,18]. For
the former, we may write $\ket{\varphi\, ,X}$ as
$$
\ket{\varphi\, ,X}=  e^{\beta\varphi(0)}{\phys}_{\rm eff},\eqno(2.10)
$$
where ${\phys}_{\rm eff}$ involves only the $X^\mu$ fields and not
$\varphi$.  (There will be no confusion between the frozen momentum $\beta$
and the spin-3 antighost $\beta$.)  The physical-state conditions (2.9)
imply that
$$
(\beta+Q)(\beta+\ft67 Q)(\beta+\ft87 Q)=0,\eqno(2.11)
$$
together with the effective physical-state conditions:
$$
\eqalign{
(\leff_0 - \deff){\phys}_{\rm eff}&=0,\cr
\leff_n{\phys}_{\rm eff}&=0,\qquad n\ge 1.\cr}\eqno(2.12)
$$
The value of the effective intercept $\deff$ is 1 when
$\beta=-\ft67 Q$ or $-\ft87 Q$, and it equals $\ft{15}{16}$ when $\beta=-Q$.
Thus these states of the $W_3$ string are described by two effective
Virasoro-string spectra, for an effective energy-momentum tensor
$\teff$ with central charge $c=\ft{51}2$ and intercepts $\deff=1$
and $\deff=\ft{15}{16}$. The first of these gives a mass
spectrum similar to an ordinary string, with a massless vector at level 1,
whilst the second gives a spectrum of purely massive states [5,6].

    As we have indicated in the introduction, the $W_3$-string spectrum is
much richer than simply that of the standard physical states we have
discussed so far. Although the classification of physical states with
non-standard ghost structure is as yet incomplete, some classes of such
states have been found [9,11,12].  They contain excitations of the ghost and
antighost fields as well as the matter fields.  The level number $\ell$ of
these states is defined with respect to the ghost vacuum $\ket{\m\m}$ given
in (2.7).  Thus, for example, the $SL(2,C)$ vacuum $\ket{0}$ has level
number $\ell=4$ and ghost number $G=-3$, since it can be written as
$\beta_{-2} \beta_{-1} b_{-1}\ket{\m\m}$.  It is straightforward to see that
at level $\ell$, the allowed ghost numbers of {\it states} (not
necessarily physical) lie in the interval
$$
1-\big[\sqrt{4\ell+1}\big] \le G \le 1+\big[\sqrt{4\ell+1}\big],
\eqno(2.13)
$$
where $\big[a\big]$ denotes the integer part of $a$.

     All the physical states in the $W_3$ string occur in multiplets [9,11].
The members of a multiplet are obtained by (repeatedly) acting with the
$G=1$ operators $a_\varphi\equiv [Q_B,\varphi]$ and $a_{X^\mu}^\dum \equiv
[Q_B,X^\mu]$ on a physical state that we call a {\it prime state}.  Until now
the known examples of prime states occurred either at ghost number $G=0$ (in
the case of states with standard ghost structure), or  at $G=-1$ or $G=-3$
(in the case of states with non-standard ghost structure). The corresponding
operators have ghost numbers $G=3$, $G=2$ or $G=0$.  All physical states in
the $W_3$ string occur with $\varphi$ momentum frozen to specific values.
Examples of physical states at level $\ell=0,1,2,3$ were given in [11] and
used for calculating scattering amplitudes.  All the prime operators of [11]
have an effective spacetime interpretation, in the sense that they can be
written as products of the form $V=U_{({\rm gh},\varphi)}\, U^{\rm eff}$,
where $U_{({\rm gh},\varphi)}$ involves only the ghosts and the $\varphi$
field, and $U^{\rm eff}$ involves only the effective spacetime coordinates
$X^\mu$.  We have already seen from (2.6) and (2.10) that states with
standard ghost structure give physical operators of this form, with $U^{\rm
eff}$ having effective intercepts $\leff_0=1$ or $\ft{15}{16}$ as measured
by $\teff$.  The prime operators with non-standard ghost structure discussed
in [11] can have three possible values for $\leff_0$, namely 1, $\ft{15}{16}$
or $\ft12$.  In section 3 we shall find new prime states at $\ell=4,\
G=-2$; $\ell=5,\ G=-2$; and $\ell=8,\ G=-3$.\footnote{$^*$}{\tenfoot In [11],
a physical state at $\ell=3$, $G=-1$ given by the operator ($A$.12) of [11] was
inadvertently described as a prime state. In fact the prime state at this
level occurs at $G=-2$, and corresponds to the operator given in ($A$.7) of
this paper; the state used in [11] is obtained by acting with the ghost
boosters $a_\varphi$ and $a_{X^\mu}^\dum$ on this prime state.} They again
all admit an effective spacetime interpretation, with the same three values
of $\leff_0$. By using these new states in scattering calculations, we shall
be able to clarify some unresolved issues in [11].  We shall also describe the
structure of the multiplets in the multi-scalar $W_3$ string.  As we shall
see, for the physical states described above, the ghost boosters $a_\varphi$
and $a_{X^\mu}^\dum$ generate a quartet from a given prime state.

     Scattering amplitudes are obtained by integrating over the worldsheet
coordinates of physical operators in correlation functions.  For
the $W_3$ string the correlation functions are, by definition, given by
functional integrals over all the quantum fields of the theory, namely the
ghosts $b$, $c$, $\beta$ and $\gamma$, the $\varphi$ field, and the
effective spacetime coordinates $X^\mu$.  These functional integrals can, as
usual, be conveniently calculated by using the techniques of conformal field
theory.

     A physical operator, denoted generically by $V(z)$, is by definition a
non-trivial BRST-invariant operator that gives rise to the physical state
$V(0)\ket{0}$.  In particular, it has conformal dimension 0, as measured by
the total energy-momentum tensor $T^{\rm tot}=T+T_{\rm gh}$.  (Note that
$V(z)$ includes ghosts as well as the matter fields.)  In an
$N$-point scattering amplitude one can use the $SL(2,C)$ invariance of
$\ket{0}$ to fix the worldsheet coordinates of three of the operators in the
correlation function, leaving $(N-3)$ coordinates over which to integrate.
For conformal covariance one must integrate over operators of dimension 1.
As in string theory [19], this can be achieved by making the replacement
$$
V(z)\rightarrow {1\over 2\pi i} \oint_z dw\, b(w)V(z),\eqno(2.14)
$$
where the subscript on the contour integral indicates that it should be
evaluated around a path enclosing $z$.  In fact this procedure not only
preserves the projective structure but also gives scattering amplitudes
that are invariant under the BRST transformations generated by (2.1) [11].
This is because the BRST variation of the right-hand side of (2.14) is a
total derivative.  Note that one cannot use the spin-3 antighost $\beta$
in place of $b$ in (2.14), since it would not give BRST-invariant amplitudes.

     In the ordinary 26-dimensional bosonic string, the physical operators
all have the form $V(z)=\big(c\, U \big)(z)$ (or its conjugate, $\big(\del
c\,c\, U \big)(z)$), where $U(z)$ involves only the coordinates $X^\mu$ but
no ghosts.   If the replacement (2.14) were not made in ($N\ge4$)-point
functions (and accordingly, the corresponding worldsheet coordinate were
left unintegrated), the results would certainly be BRST invariant.  However,
they would trivially be zero, since the non-zero ghost inner product in
string theory is $\bra{0} \del^2 c\, \del c\, c\ket{0}$ [19].  In the $W_3$
string, on the other hand, physical operators do not all have the standard
ghost structure, so it is no longer {\it a priori} obvious that the
replacement (2.14) is required in order to get non-zero results for
BRST-invariant $(N\ge4)$-point scattering amplitudes. Nevertheless, it seems
that the corresponding correlation functions in the $W_3$ string vanish
unless the replacement (2.14) is made, as we shall discuss in section 6.
(The situation is different if physical states with discrete
effective-spacetime momentum are involved; see section 4.)

     In order to identify which correlation functions might be non-vanishing,
it is useful first to consider two necessary conditions [11].  If these
conditions are satisfied, then it becomes a matter of more detailed
calculation to determine the result.  The first of the necessary conditions
is that the total ghost structure of the product of operators in the
correlation function should be appropriate.  Specifically, the non-vanishing
ghost inner product in the $W_3$ string is given by
$$
1=\bra{0} c_{-1}c_0c_1\,\gamma_{-2}\gamma_{-1}\gamma_0
\gamma_1\gamma_2\ket{0}
=\ft1{576}\bra{0} \del^2 c\,\del c \,c\,
\del^4\gamma\,\del^3\gamma\,\del^2\gamma\,\del\gamma\,\gamma\ket{0}\ .
\eqno(2.15)
$$
(See [11] for a more detailed discussion.)  Note that in particular the total
ghost number of the operators in a non-vanishing correlator must be $3+5=8$.

     The second of the necessary conditions for obtaining a non-vanishing
correlation function is that momentum conservation must be satisfied.
Owing to the presence of the background charges, this implies that we must
have $\sum_{i=1}^N p^\mu_i =-2a^\mu$ in the effective spacetime, together with
$$
\sum_{i=1}^N \beta_i=-2Q \eqno(2.16)
$$
in the $\varphi$ direction.  For states of continuous spacetime momentum
$p_\mu$, as indeed we have in the multi-scalar $W_3$ string,  momentum
conservation in the $X^\mu$ directions can be straightforwardly  satisfied.
However, the momentum $\beta$ in the $\varphi$ direction can only take
specific frozen values in physical states. (For physical states with
standard ghost structure this follows from (2.11); it is true also for
states with non-standard ghost structure, as we shall see again in the
next section.)  Thus it is in general non-trivial to satisfy momentum
conservation in the  $\varphi$ direction.  Note that (2.16) is an extremely
restrictive condition; indeed it implies that $(N\ge 3)$-point correlation
functions built exclusively from physical operators of standard ghost
structure will vanish, as may be seen from (2.11).  It is worth emphasising
that even though $\varphi$ does not admit the interpretation of being an
observable spacetime coordinate, the functional integration over it, and in
particular over its zero-mode, determines a very stringent $W_3$ selection
rule [11].

     Using the procedure outlined above, we calculated various three-point
and four-point functions in [11], with the physical states given therein.  A
number of patterns emerged, which we shall be able to develop further with
the results of the present paper.  In particular, it was found that
physical states seem to be characterised by their effective spacetime
structure, in the sense that physical states with the same $\leff_0$ values
but different ghost and $\varphi$ dependence give identical results in
correlation functions, modulo an overall constant factor.  This factor
depends on the normalisation of the states, and furthermore would be zero
if, for example, a selection rule such as (2.16) forbade a particular
interaction.

     It was found in [11] that non-vanishing three-point functions can occur
for physical operators with $\leff_0$ values of $\{1,\ 1,\ 1\}$,
$\{\ft{15}{16},\ \ft{15}{16},\ 1\}$, $\{\ft12,\ \ft12,\ 1\}$ and
$\{\ft{15}{16},\ \ft{15}{16},\ \ft12\}$.  All other three-point functions
vanish.  These results for three-point functions are in one-to-one
correspondence with the fusion rules of the two-dimensional Ising model, if
one associates the $\leff_0=1$ sector with the identity operator ${\bf 1}$,
the $\leff_0=\ft{15}{16}$ sector with the spin operator $\sigma$, and the
$\leff_0=\ft12$ sector with the energy operator $\varepsilon$.  This makes
more precise the numerological connection between $W_3$ strings and the
Ising model which emerges from the analysis of the spectrum [3,6,10].  In
particular, the central charge of the effective energy-momentum tensor is
$\ft{51}2=26-\ft12$, where 26 is the critical central charge of the usual
bosonic string, and $\ft12$ is the central charge of the Ising model;
moreover, the set of $\leff_0$ values $\{1,\ \ft{15}{16},\ \ft12\}$ can be
written as $1-\Delta_{\rm min}$, where 1 is the intercept of the usual
bosonic string and $\Delta_{\rm min}$ takes the values of the dimensions of
the primary fields $\{{\bf 1},\ \sigma,\ \varepsilon\}$ of the Ising model.

     One of the interesting results of [11] is that this connection between
the $W_3$ string and the Ising model does not go both ways in higher-point
functions.  In particular, it was shown in [11] that the selection rule (2.16)
forbids the existence of a four-point function for four physical operators
with $\leff_0=\ft{15}{16}$, even though the four-point function
$\fpf\sigma\sigma\sigma\sigma$ is non-vanishing in the Ising
model.  We shall find further indications in section 4 from five-point and
six-point functions that the $\leff_0=\ft{15}{16}$ sector plays a special
r\^ole in the $W_3$ string.  One of the general features that we observed in
[11] is that four-point functions in the $W_3$ string exhibit a duality
property, which means that they can be interpreted in terms of underlying
three-point functions, with sets of intermediate states being exchanged in
the $s$, $t$ or $u$ channels. In this paper we shall illustrate this
factorisation property for a five-point function too.

\bigskip
\bigskip
\noindent{\bf 3. New physical states and the multiplet structure}
\bigskip

     In this section, we shall construct some new physical states of the
$W_3$ string with non-standard ghost structure.  These are prime states at
levels $\ell=4$ and 5, with ghost number $G=-2$.  We shall use these, and an
$\ell=8$, $G=-3$ prime state which was found in [12], to calculate more
scattering amplitudes in the next section.  In particular, the $\ell=5$
prime state enables us to obtain a non-vanishing four-point function with
four $\leff_0=1$ operators, which does not exist for the states given in [11].
The above states will provide more evidence for the observation that
physical states appear to be characterised by their effective spacetime
structures.  We shall also investigate the multiplet structures generated by
the action of the ghost boosters $a_\varphi$ and $a_{X^\mu}^\dum$.

     Until now, prime states have been discovered at ghost number $G=0$ for
states with standard ghost structure, and at $G=-1$ and $G=-3$ in the case
of states with non-standard ghost structure.  The two new prime states that
we shall construct in this section occur at $G=-2$, and thus are of a
different kind.

     The general procedure that we use for finding physical states is as
follows.  At a given, fixed, level number $\ell$ and ghost number $G$ we
write down the most general state, as a sum of all $n_G$ possible structures
with arbitrary coefficients $g_i$.  Requiring BRST invariance leads to
$n_{G+1}$ equations for the $g_i$'s, giving $m_G$ independent solutions.
There are $n_{G-1}-m_{G-1}$ BRST-trivial states at ghost number $G$, where
$m_{G-1}$ denotes the number of BRST-invariant states at ghost number $G-1$.
The number of BRST-nontrivial physical states at ghost number $G$ is
therefore given by $m_G-(n_{G-1}-m_{G-1})$.  Thus at a given level number
$\ell$ we may start from the lowest ghost number given by (2.13), and
systematically work up to the highest ghost number allowed by (2.13),
finding all the non-trivial physical states at that level.  This is quite a
tedious process, which is best performed with the aid of a computer.

    Applying this procedure at level 4, we have found a physical state at
$G=-2$, with $\varphi$ momentum $\beta=\ft17Q$.  The general solution has
three parameters which means, since there are two BRST-trivial structures
coming from $Q_B$ acting on the two structures at $G=-3$, that there is one
non-trivial solution at $G=-2$.  The two trivial parameters may be used in
order to give the physical state a convenient form.  In fact, it turns out
that they may be chosen so as to remove all terms involving excitations of
the spacetime coordinates, $X^\mu$, and thus the state acquires an effective
spacetime interpretation, as a tachyon.  The explicit form of the
corresponding physical operator, which has ghost number $G=1$, is given in
the appendix in ($A$.8).  For the amplitudes that we shall be concerned with
in this paper, only two of its thirteen terms contribute.  They are
$$
\V{1}{15/16}{\ft17Q}{p} =-\ft3{16}\sqrt{29}\,i\, \Big(3\sqrt{2}\, c\,
\beta\, \gamma  +4 \del\varphi\, c+ \cdots \Big)
e^{\ft17 Q\varphi} e^{ip\cdot X}\ ,\eqno(3.1)
$$
where the effective spacetime momentum $p_\mu$ satisfies the mass-shell
condition
$$
\ft12 p^\mu(p_\mu +2 a_\mu)=\ft{15}{16}\ .\eqno(3.2)
$$
In (3.1) we have used the notation of [11], where a physical operator $V$ with
ghost number $G$, $\leff_0=\Delta$ and momentum $(\beta,p_\mu)$ is denoted
by $\V{G}{\Delta}{\beta}{p}$.  The operator (3.1) is a representative of
a tachyon of the $\leff_0=\ft{15}{16}$ sector of the $W_3$ string.  The
operator $e^{ip\cdot X}$ may be replaced by an arbitrary excited effective
physical operator $R({\scriptstyle\del X^\mu})e^{i p' \cdot X}$ with
the same $\leff_0$ value.  This allows the construction of new physical
states at higher levels with $X^\mu$ excitations, that have an
interpretation as excited effective spacetime states.

     At level 5, we have found another new physical state at $G=-2$, with
$\varphi$ momentum $\beta=\ft27Q$.  Again we can use the freedom to add
BRST-trivial terms so as to give the state an effective spacetime
interpretation, as a tachyon.  The corresponding physical operator is given
in ($A$.9).  For the amplitudes that we shall be concerned with in this paper,
only 5 of its 32 terms contribute, namely
$$
\eqalign{
\V{1}{1}{\ft27Q}{p} = \ft1{20}\sqrt{29}\,i\, \Big( &6\sqrt{2}\, c\,\del
\beta\,\gamma  -3\sqrt{2}\, c\,\beta\,\del\gamma+
12 \del\varphi\, c\, \beta\,\gamma\cr
& + 4\sqrt{2}\, \del\varphi\,\del\varphi\, c
+ 2 \del^2\varphi\, c +\cdots \Big)e^{\ft27 Q\varphi}
e^{ip\cdot X}\ ,\cr}\eqno(3.3)
$$
where the effective spacetime momentum $p_\mu$ satisfies the mass-shell
condition
$$
\ft12 p^\mu (p_\mu + 2 a_\mu) = 1\ .\eqno(3.4)
$$
This operator is a representative of an effective tachyon of the
$\leff_0=1$ sector. Again, one can replace $e^{i p\cdot X}$ by an arbitrary
excited effective-spacetime physical operator with the same $\leff_0$ value.

     The third physical state that we shall be using in this paper, and that
was not given in [11], occurs at level 8 with ghost number $G=-3$ and
$\varphi$ momentum $\beta=\ft47Q$; it was found in [12]. Once more, the
freedom of adding BRST trivial parts can be used to remove all excitations
in the $X^\mu$ directions, thus giving it the interpretation of an effective
spacetime tachyon. As such it is a representative of the $\leff_0=\ft12$
sector.  It has 82 terms, and is given in ($A$.10).  In fact, for the
scattering amplitudes that we shall calculate in the next section, only four
of those terms contribute to the result, namely
$$
\V{0}{1/2}{\ft47Q}{p}= \Big( \del\varphi\, c\,\del\beta + 3\sqrt{2}\,
c\,\del\beta\,\beta\,\gamma - 3 \del^2\varphi\, c\,\beta -\ft34\sqrt{2}\,
c\,\del^2\beta +\cdots \Big)e^{\ft47Q\varphi} e^{ip\cdot X}\ ,\eqno(3.5)
$$
where
$$
\ft12 p^\mu (p_\mu + 2 a_\mu) = \ft12\ .\eqno(3.6)
$$

     The fact that only the small subsets of terms (3.1), (3.3) and (3.5)
contribute in scattering amplitudes involving the physical operators ($A$.8),
($A$.9) and ($A$.10) respectively suggests that there is a great deal of
redundancy in the formalism describing the $W_3$ string.  In section 6, we
shall introduce a new formalism that removes this redundancy.

     The new physical operators that we have presented here have lower ghost
numbers than any of those used in [11].  Specifically, ($A$.8) and ($A$.9) have
$G=1$, and ($A$.10) has $G=0$.  An important consequence of this is that it
enables us to obtain non-vanishing $N$-point functions for arbitrarily large
$N$, whereas the physical operators of [11] only allowed their construction
for $N\le 5$.  This is because the total ghost number of operators in a
non-vanishing $N$-point function must be equal to 8, and the insertion of
many operators of the form (2.14) will rapidly cause this number to be
exceeded unless the physical operators $V$ have ghost numbers $\le1$.  We
shall discuss this further at the end of section 4.  This completes our
discussion of new prime states of the $W_3$ string with continuous on-shell
momenta.

     Let us now turn our attention to the structure of the multiplets
generated by the action of the ghost boosters $a_\varphi$ and
$a_{X^\mu}^\dum$ on prime states. The explicit form of the ghost boosters is
given in [9] for the two-scalar $W_3$ string and the generalisation to the
multi-scalar case is immediate.  They are given in an alternative
formulation in (6.21).  Although we do not have a general proof, we have
checked the structure of the multiplets in several examples.  Acting on a
prime state $\ket{G}$ at ghost number $G$ with the $(d+1)$ ghost boosters,
we obtain just two independent non-trivial physical states at ghost number
$(G+1)$: $a_\varphi\ket{G}=\ket{G+1}_1$ and $a_{X^\mu}^\dum\ket{G}=
(p^\mu+a^\mu)\ket{G+1}_2$. The fact that $a_{X^\mu}^\dum\ket{G}$ gives only
one non-trivial state rather than $d$ is quite surprising and results in a
much simpler multiplet structure than one might have expected (we shall
return to this point in section 6). The multiplet is completed by a single
non-trivial state $\ket{G+2}$ obtained by acting either with $a_\varphi$ on
$\ket{G+1}_2$ or with $a_{X^\mu}^\dum$ on $\ket{G+1}_1$. Thus it seems from
our examples that each prime state $\ket{G}$ gives rise to a quartet of
physical states $\{ \ket{G},\ \ket{G+1}_1,\ \ket{G+1}_2,\ \ket{G+2}\}$.  The
structure is reminiscent of an $N=2$ supermultiplet.

     As discussed in [9,11], each physical state at ghost number $G$ and
momentum $(\beta,p_\mu)$ has a conjugate partner at ghost number $(2-G)$ and
momentum $(-2Q-\beta,-2a_\mu-p_\mu)$.  Thus associated with each prime state
is a quartet and a conjugate quartet of physical states.

     We have already remarked that for all the known examples, prime states
either automatically have, or can be given, an effective spacetime
interpretation. Remarkably, in all the examples we have checked, the freedom
to add BRST-trivial parts is precisely sufficient to enable the boosted
members of a multiplet to acquire an effective spacetime interpretation too.
In one of our examples, the prime state involved an excited effective
spacetime part, indicating that this effective spacetime interpretation is
universal, and not just confined to effective tachyonic states.

     These considerations lead us to conjecture that {\it all} the physical
states (with continuous on-shell spacetime momentum $p_\mu$ --- see below) in
the $W_3$ string admit an effective spacetime interpretation, {\it i.e.}\
they can all be written in the form
$$
\ket{{\rm ghost}+\varphi}\otimes \ket{{\rm effective\ spacetime}}
\ .\eqno(3.7)
$$
We shall sharpen this conjecture in section 6.  Moreover, we conjecture that
the effective intercepts of all these physical states take values in the set
$\{1,\ \ft{15}{16},\ \ft12\}$.\footnote{$^*$}{\tenfoot In [13], physical
states at $\ell=2$, $G=2$ in the two-scalar $W_3$ string were discussed,
corresponding to an operator of the form $\del^3 c\, c\, \del^2\gamma\,
\del\gamma\, \gamma e^{\beta_1\varphi_1} e^{\beta_2 \varphi_2}$. This is
annihilated by $Q_B$ provided that the exponentials have total conformal
weight 2.  However, contrary to what is asserted in [13], this state is in
general BRST trivial, except for three values of $(\beta_1,\beta_2)$. Two of
these values, $(-\ft{12}7Q, -\ft{12}7a)$ and $(-\ft{12}7Q, -\ft{2}7a)$,
correspond to states with $\leff_0=\ft12$ which generalise to an
$\leff_0=\ft12$ state in the multi-scalar $W_3$ string. (In fact this is the
conjugate of one of the quartet members from the prime state given by
$\V{2}{1/2}{-\ft27Q}{p}$ in formula ($A$.6).) The third momentum
value, $(-Q,-\ft{17}7 a)$, corresponds to a state with
$\leff_0=-\ft{17}{16}$ that does not generalise to the multi-scalar $W_3$
string.  Our conjecture that the effective intercept values are $\{1,\
\ft{15}{16},\ \ft12\}$ relates to the physical states of the {\it
multi}-scalar $W_3$ string, and thus the $\leff_0=-\ft{17}{16}$ state does
not contradict it.}

     The above discussion of the effective-spacetime interpretation of
physical states, and their multiplet structure, applies for all physical states
with continuous on-shell spacetime momentum $p_\mu$.  There are, however,
additional physical states in the spectrum of the multi-scalar $W_3$ string
that occur only for fixed values of $p_\mu$.  In fact, in all the examples
we know, these discrete states occur with $p_\mu=0$.  The simplest example
is provided by the $SL(2,C)$ vacuum, $\ket{0}=\beta_{-2}\beta_{-1} b_{-1}
\ket{\m\m}$, which is a level $\ell=4$ state with ghost number $G=-3$.  We
shall adopt the notation $\D{G}{\beta}$ to denote the physical operator with
ghost number $G$ and $\varphi$ momentum $\beta$ that gives a discrete state
with ghost number $(G-3)$.  Thus $\ket{0}$ corresponds to the operator
$\D{0}{0}=1$.

     We shall first consider the prime discrete states, and discuss their
multiplet structures later.  There are two examples at level $\ell=1$,
corresponding to the following operators [9]:
$$
\eqalignno{
\D{2}{-\ft67Q}&= \big(c\, \gamma -{i\over 3\sqrt{58}} \del\gamma\, \gamma
\big) e^{-\ft67 Q\varphi}\ ,&(3.8)\cr
\D{2}{-\ft87Q}&= \big(c\, \gamma +{i\over 3\sqrt{58}} \del\gamma\, \gamma
\big) e^{-\ft87 Q\varphi}\ .&(3.9)\cr}
$$
Apart from the $SL(2,C)$ vacuum, the next example of a prime discrete state
that we know of occurs at level $\ell=6$, ghost number $G=-3$ [9], with the
corresponding operator being denoted by $\D{0}{\ft27Q}$.  (In the two-scalar
$W_3$ string [9], it plays a similar r\^ole to the ground-ring generators of
Witten [15].)  From its detailed form, which is given explicitly in [9] for the
two-scalar $W_3$ string, one can see that the freedom to add BRST-trivial
parts is precisely sufficient to enable it to be given an effective spacetime
interpretation; the result appears in ($A$.14).  This is presumably the
generic situation for discrete prime states in the multi-scalar $W_3$ string.
Higher-level discrete states can be generated by taking appropriate
normal-ordered products of these examples.  For instance,
$\big(\D{0}{\ft27}\big)^{4m}$ gives the new discrete operator $\D{0}{\ft87 m
Q}$ at level $\ell=(4m^2 + 7m+4)$, and $\big(\D{0}{\ft27}\big)^{4m+1}$ gives
the new discrete operator $\D{0}{\ft27(4m+1)Q}$ at level $\ell=(4m^2 +11m
+6)$ [9]. Thus we have discrete operators with $G=0$ and
$$
\eqalign{
\D0{\ft87 mQ} &: \qquad \ell=4 m^2+7 m+4\ ,\cr
\D0{\ft27 (4m+1)Q} &: \qquad \ell=4 m^2+11 m+6\ .\cr}\eqno(3.10)
$$

     We now turn to the consideration of the multiplet structure associated
with the prime discrete states.  Here we find by checking several
examples that the pattern is rather different from that for the
continuous-momentum physical states; for $d$ effective spacetime coordinates
$X^\mu$ we get a multiplet with $4d$ members.  It is convenient to decompose
the ghost boosters $a_{\scriptscriptstyle X^\mu}^\dum$ into an operator
$a^{\scriptscriptstyle /\!/}$ which is parallel to the background charge
vector $a^\mu$ and $(d-1)$ operators $(a_{\scriptscriptstyle X^\mu})^\perp$
which are perpendicular to the background charge vector. Acting on a prime
discrete state $\ket{G}$ of ghost number $G$, we have
$$
\eqalign{
G&: \qquad\qquad\qquad\qquad\qquad\ket{G}\cr
G+1&: \quad\qquad a_\varphi\ket{G},\quad\qquad a^{\scriptscriptstyle /\!/}
  \ket{G}, \quad\qquad
      (a_{\scriptscriptstyle X^\mu})^\perp\ket{G}\cr
G+2&: \qquad a_\varphi\,a^{\scriptscriptstyle /\!/}\ket{G},\quad
a_\varphi\,
   (a_{\scriptscriptstyle X^\mu})^\perp\ket{G},
 \quad a^{\scriptscriptstyle /\!/} \,
  (a_{\scriptscriptstyle X^\mu})^\perp  \ket{G}
 \cr
G+3&: \qquad\qquad\qquad\quad a_\varphi\, a^{\scriptscriptstyle /\!/}\,
     (a_{\scriptscriptstyle X^\mu})^\perp\ket{G}\ .\cr}\eqno(3.11)
$$
\np
Thus we see that the multiplicities of states at ghost numbers $\{G,\ G+1,\
G+2,\ G+3\}$ are $\{1,\ d+1,\ 2d-1,\ d-1\}$.  The situation is very
different from the case of prime states with continuous momentum, in that
here each of the $(d+1)$ ghost boosters generates an independent non-trivial
physical state at ghost number $(G+1)$. Note that the repeated application
of the $(a_{\scriptscriptstyle X^\mu})^\perp$ operators does not give any
non-trivial state, so (3.11) gives the complete multiplet. In some sense it
is reminiscent of an $N=3$ supermultiplet.\footnote{$^*$}{\tenfoot It seems
that a natural generalisation of the decomposition of $a_{\scriptscriptstyle
X^\mu}^\dum$ into $a^{\scriptscriptstyle /\!/}$ and $(a_{\scriptscriptstyle
X^\mu})^\perp$ for physical states with continuous momentum is to decompose
$a_{\scriptscriptstyle X^\mu}^\dum$ into its components parallel and
perpendicular to $(p_\mu+a_\mu)$. In our discussion of the quartet structure
of the multiplets for continuous-momentum physical states, we saw that when
$a_{\scriptscriptstyle X^\mu}^\dum$ acts on such physical states, it gives
states of the form $(p_\mu+a_\mu)\ket{\dum}$. Thus $(a_{\scriptscriptstyle
X^\mu})^\perp$ annihilates continuous-momentum physical states, and we see
that the multiplet structure in (3.11) reduces to a quartet.}

     Another striking difference between the discrete states and the
physical states with continuous momentum $p_\mu$ concerns their effective
spacetime interpretation.  First, we recall that {\it all} prime physical
states, discrete or continuous, appear to admit an effective spacetime
interpretation.  For continuous-momentum physical states, this is true also
for the entire multiplet ({\it i.e.}\ quartet) generated by the ghost boosters.
For the discrete states, however, this is no longer the case; the multiplet
members do not all admit an effective-spacetime interpretation, even though
the prime state does.  Specifically, it is states generated by the action of
$a_{X^\mu}^\dum$ that lack an effective-spacetime interpretation.  The
reason for this is that the Lorentz index ${\scriptstyle \mu}$ has to live
either on the background-charge vector $a^\mu$ or on the effective-spacetime
coordinates $X^\mu$.  If it lived solely on $a^\mu$, then the
$a_{X^\mu}^\dum$ operators would give only one independent physical state,
whereas in fact they all give independent states in the multiplet.  Thus in
some (but not, as can easily be checked, all) of the terms in the boosted
states the ${\scriptstyle \mu}$ index must live on $X^\mu$, implying
that these states cannot all be given an effective-spacetime interpretation.

     The discrete operators can be used to map physical states with
continuous momentum $p_\mu$ into others by normal ordering, provided that
the normal ordered product exists. This appears to be their most important
r\^ole in the interacting $W_3$ string, since if one tries to view them as
ordinary physical states in their own right, they give divergent results in
$(N\ge 4)$-point amplitudes. We shall discuss this further in the next
section.

     For convenience, we summarise the prime operators that we shall be using
in this paper in two tables.  Their explicit forms are given in the
appendix. (In Table 1 we have included a physical operator at level $\ell=9$,
which we shall find  in section 6 in a new formalism.)
\np
$$
\hbox{
\vbox{\tabskip=0pt \offinterlineskip
\def\tablerule{\noalign{\hrule}}
\halign to250pt{\strut#& \vrule#\tabskip=0em plus10em&
\hfil#\hfil& \vrule#& \hfil#\hfil& \vrule#& \hfil#\hfil& \vrule#&
\hfil#& \vrule#\tabskip=0pt\cr\tablerule
&&\phantom{}&&${\scriptstyle G}$&&${\scriptstyle L_0^{\rm eff}}$
&&${\scriptstyle \beta\ \qquad}$&\cr\tablerule
&&\phantom{}&&${\scriptstyle 3}\,$&&${\scriptstyle 15/16 }$
&&${\scriptstyle -Q}\ \qquad$&\cr
&&$\ell=0$
&&${\scriptstyle 3}$
&&${\scriptstyle 1}$
&&${\scriptstyle -6Q/7, \quad -8Q/7}$&\cr
\tablerule
&&\phantom{}&&${\scriptstyle 2}\,$&&${\scriptstyle 15/16 }$
&&${\scriptstyle -3Q/7}$\qquad&\cr
&&$\ell=1$
&&${\scriptstyle 2}$
&&${\scriptstyle1/2}$
&&${\scriptstyle -4Q/7}\qquad$&\cr
\tablerule
&&$\ell=2$
&&${\scriptstyle 2}\,$&&${\scriptstyle 1/2}$
&&${\scriptstyle -2Q/7}\qquad$&\cr \tablerule
&&$\ell=3$
&&${\scriptstyle 1}\,$&&${\scriptstyle 1}$&&${\scriptstyle 0}\qquad\quad$&\cr
\tablerule
&&$\ell=4$
&&${\scriptstyle 1}\,$&&${\scriptstyle 15/16}$&&${\scriptstyle  Q/7}
\qquad$&\cr
\tablerule
&&$\ell=5$
&&${\scriptstyle 1}\,$&&${\scriptstyle 1}$&&${\scriptstyle 2Q/7}
\qquad$&\cr
\tablerule
&&$\ell=8$
&&${\scriptstyle 0}\,$&&${\scriptstyle 1/2}$&&${\scriptstyle 4Q/7}
\qquad$&\cr
\tablerule
&&$\ell=9$
&&${\scriptstyle 0}\,$&&${\scriptstyle 15/16}$&&${\scriptstyle 5Q/7}
\qquad$&\cr
\tablerule
\noalign{\bigskip}}}}
$$
\centerline{\it Table 1. Continuous-momentum physical operators}

\bigskip
$$
\hbox{
\vbox{\tabskip=0pt \offinterlineskip
\def\tablerule{\noalign{\hrule}}
\halign to250pt{\strut#& \vrule#\tabskip=0em plus10em&
\hfil#\hfil& \vrule#& \hfil#\hfil& \vrule#&
\hfil#& \vrule#\tabskip=0pt\cr\tablerule
&&\phantom{}&&${\scriptstyle G}$
&&${\scriptstyle \beta\ \qquad}$&\cr\tablerule
&&$\ell=1$
&&${\scriptstyle 2}$
&&${\scriptstyle -6Q/7, \quad -8Q/7}$&\cr
\tablerule
&&$\ell=4$
&&${\scriptstyle 0}\,$
&&${\scriptstyle 0}\qquad\quad$&\cr \tablerule
&&$\ell=6$
&&${\scriptstyle 0}\,$
&&${\scriptstyle 2Q/7}\qquad$&\cr \tablerule
\noalign{\bigskip}}}}
$$
\centerline{\it Table 2. Discrete physical operators}
\bigskip

\bigskip
\bigskip
\noindent{\bf 4. $W_3$-string scattering}
\bigskip

       In our previous paper [11] on the interacting $W_3$ string, we evaluated
all the non-vanishing four-point functions for effective tachyons that could
be built at tree level using the physical states given in that paper.  As we
have already remarked, they are characterised by their effective-spacetime
structure, and in particular their $\leff_0$ values. For convenience, we
list the structure of the results here. In an obvious notation, we have
[11]:
$$
\eqalignno{
\fpf{\ft12}{\ft12}{\ft12}{\ft12}&\sim \int_0^1 dx\ x^{-s/2-2}
                                 (1-x)^{-t/2-2}(1-x+x^2)\ ,
&(4.1a)\cr
\fpf{\ft{15}{16}}{\ft{15}{16}}{\ft12}{\ft12}&\sim \int_0^1 dx\ x^{-s/2-2}
                                 (1-x)^{-t/2-31/16}(x-2)\ ,
&(4.1b)\cr
\fpf{1}{\ft{15}{16}}{\ft{15}{16}}{1}&\sim \int_0^1 dx\ x^{-s/2-31/16}
                                 (1-x)^{-t/2-2}\ ,
&(4.1c)\cr
\fpf{\ft{15}{16}}{\ft{15}{16}}{1}{\ft12}&\sim \int_0^1 dx\ x^{-s/2-3/2}
                                 (1-x)^{-t/2-31/16}\ ,
&(4.1d)\cr
\fpf{1}{\ft12}{1}{\ft12}&\sim \int_0^1 dx\ x^{-s/2-3/2}
                                 (1-x)^{-t/2-3/2}\ ,
&(4.1e)\cr}
$$
where $s$, $t$ (and $u$) are the Mandelstam variables:
$$
\eqalign{
s&\equiv -(p_1+p_2)^2 - 2 a \cdot (p_1+p_2)\ ,\cr
t&\equiv -(p_2+p_3)^2 - 2 a \cdot (p_2+p_3)\ ,\cr
u&\equiv -(p_1+p_3)^2 - 2 a \cdot (p_1+p_3)\ .\cr}\eqno(4.2)
$$
In [11] we observed that the above four-point functions have the correct
crossing properties, and that they exhibit  a $W_3$ duality behaviour in
which they can be written in terms of the underlying three-point functions
with the exchange of intermediate states in the $s$, $t$ or $u$ channels.
{}From the underlying three-point functions, one might expect that there could
be two more non-vanishing four-point functions, namely, $\fpf1111$ and
$\fpf{\ft{15}{16}}{\ft{15}{16}}{\ft{15}{16}}{\ft{15}{16}}$.  For the latter, we
proved in [11] that there cannot exist $\leff_0=\ft{15}{16}$ physical states
with $\varphi$ momenta that enable the selection rule (2.16) to be satisfied
in this case. In other words
$$
\fpf{\ft{15}{16}}{\ft{15}{16}}{\ft{15}{16}}{\ft{15}{16}} =0\eqno(4.3)
$$
in the $W_3$ string. Later in this section we shall argue that in fact any
$N$-point function with three or more $\leff_0=\ft{15}{16}$ physical states
must vanish.

      For the case of $\fpf1111$ on the other hand, we argued in [11] that
there was no reason to expect that it must vanish, and that our inability to
construct it was simply due to an insufficient supply of examples of
physical states with $\leff_0=1$. In fact with the new physical states given
in this paper, we shall indeed be able to construct such a four-point
function.  The other new physical states given in section 3 enable us to
construct further examples of four-point functions that provide additional
evidence for the conjecture that they are characterised by the
effective-spacetime structure of the physical states.

     We begin the detailed discussion of our new results by considering the
four-point function $\fpf1111$ mentioned above.  This can be constructed,
for example, by taking the level $\ell=5$ operator $\V11{\ft27 Q}{p}$ given
in ($A$.9) together with physical operators given in [11]. One of them is the
level $\ell=0$ operator $\V31{-\ft87 Q}{p}$ of standard ghost structure (of
which we take two copies); the remaining operator, $\aV210p$, has level
number $\ell=3$.  It is obtained from $\V110{p}$ given in ($A$.7) by
boosting; its explicit form is given in ($A$.12) of [11], where it was
inadvertently denoted as $\V210{p}$.  We choose to make the replacement
(2.14) on the level $\ell=5$ operator.  This replacement singles out all the
terms involving an undifferentiated $c$ ghost, and removes it, reducing the
32 terms in ($A$.9) to 14. Of these 14 terms, only 5 can give rise to the
ghost structure (2.15), namely those that do not involve a $b$ antighost or
any of its derivatives; they are given in (3.3). The $\V31{-\ft87 Q}{p_1}$
and $\V31{-\ft87 Q}{p_2}$ operators each have only one term, and 7 of the 11
terms of $\aV210{p_4}$ do not contribute. Thus there are $5\times 4=20$
different contributions to the final result. The techniques used for
calculating these contributions  were explained in [11]. Since the
intermediate steps are rather involved, we shall only present the final
result. Explicitly we find for this four-point function:
\medskip
\noindent{ $\bullet\ \underline{L^{\rm eff}_0=\{1,\ 1,\ 1,\ 1\}}$:}
$$
\eqalign{\int\oint_{z_3}
\bra{0} &\V{3}{1}{-\ft87 Q}{p_1}(z_1)\; \V{3}{1}{-\ft87 Q}{p_2} (z_2)\,
b(w)\,\V{1}{1}{\ft27 Q}{p_3}(z_3)\;  \aV{2}{1}{0}{p_4}(z_4)\ket{0}\cr
&=-{\sqrt{58}\, i\over 10}\, \int_0^1 \! dx \, x^{-s/2-2}
(1-x)^{-t/2-2}\ ,\cr}\eqno(4.4)
$$
where the integrals at the front of the first line denote the integration
over the worldsheet coordinate $z_3$, and the contour integral of (2.14),
respectively. The expression (4.4) is identical (up to normalisation) to the
four-point tachyon scattering amplitude in ordinary bosonic string theory.
This is not surprising, in view of our observation that interactions in the
$W_3$ string are characterised by the effective spacetime structure, and in
particular the $\leff_0$ values, of the external states, and that the
$\leff_0=1$ sector of the $W_3$ string has the closest resemblance to the
usual bosonic string. Indeed the result provides a reassuring check that
despite the complexity of the computation, the expected result emerges.  In
section 6, we shall present a new formalism within which it will become
manifest that tree-level scattering of purely $\leff_0=1$ states in the
$W_3$ string is the same as in the bosonic string, apart from the presence
of the background charge.

     The physical operators ($A$.8) and ($A$.10) provide further nice checks of
the observation that what characterises an interaction is the effective
spacetime structure of the external states. Using them, we have computed the
following four-point functions:
\medskip
\noindent{ $\bullet\ \underline{L^{\rm eff}_0=\{1,\ \ft{15}{16},\
\ft{15}{16},\ 1\}}$:}
$$
\eqalign{\int\oint_{z_3}
\bra{0} &\V{3}{1}{-\ft67 Q}{p_1}(z_1)\; \V{2}{15/16}{-\ft37 Q}{p_2} (z_2)\,
b(w)\,\V{1}{15/16}{\ft17 Q}{p_3}(z_3)\;  \V{3}{1}{-\ft67 Q}{p_4}(z_4)\ket{0}\cr
&=0\ ,\cr}\eqno(4.5)
$$
\medskip
\noindent{ $\bullet\ \underline{L^{\rm eff}_0=\{\ft{15}{16},\ 1,\ \ft12,\
\ft{15}{16}\}}$:}
$$
\eqalign{\int\oint_{z_3}
\bra{0} &\V{3}{15/16}{- Q}{p_1}(z_1)\; \V{3}{1}{-\ft67 Q}{p_2} (z_2)\,
b(w)\,\V{2}{1/2}{-\ft27 Q}{p_3}(z_3)\;  \V{1}{15/16}{\ft17 Q}{p_4}(z_4)
\ket{0}\cr
&=0\ ,\cr}\eqno(4.6)
$$
\medskip
\noindent{ $\bullet\ \underline{L^{\rm eff}_0=\{\ft12,\ \ft{15}{16},\
\ft{15}{16},\ \ft12\}}$:}
$$
\eqalign{\int\oint_{z_3}
\bra{0} &\V{2}{1/2}{-\ft47 Q}{p_1}(z_1)\; \V{3}{15/16}{- Q}{p_2} (z_2)\,
b(w)\,\V{1}{15/16}{\ft17 Q}{p_3}(z_3)\;  \aV{3}{1/2}{-\ft47 Q}{p_4}(z_4)
\ket{0}\cr
&=-{3\sqrt{58}\, i\over 8} \, \int_0^1 \! dx \, x^{-s/2-31/16}
(1-x)^{-t/2-2}(1+x)\ ,\cr}\eqno(4.7)
$$
\medskip
\noindent{ $\bullet\ \underline{L^{\rm eff}_0=\{1,\ 1,\ \ft12,\ \ft12\}}$:}
$$
\eqalign{\int\oint_{z_3}
\bra{0} &\aV{4}{1}{-\ft87 Q}{p_1}(z_1)\; \V{3}{1}{-\ft87 Q}{p_2} (z_2)\,
b(w)\,\V{0}{1/2}{\ft47 Q}{p_3}(z_3)\;  \V{2}{1/2}{-\ft27 Q}{p_4}(z_4)
\ket{0}\cr
&=-{\sqrt{2}\, \over 4} \, \int_0^1 \! dx \, x^{-s/2-2}
(1-x)^{-t/2-3/2}\ .\cr}\eqno(4.8)
$$
As in [11], we use the notation $\aV{G}{\Delta}{\beta}{p}$ for a physical
operator which is obtained by acting once with the ghost boosters
$a_\varphi$ or $a_{X^\mu}^\dum$ on the prime operator
$\V{G-1}{\Delta}{\beta}{p}$.\footnote{$^*$}{\tenfoot Since, as we have
discussed in section 3, there are two independent such boosted operators for
a prime operator with continuous momentum $p_\mu$, the choice of boosted
operator $\aV{G}\Delta\beta{p}$ might seem ambiguous since an arbitrary
linear combination could be taken.  However, as we have
seen, only the coefficient, but not the form, of the scattering amplitude
should depend on which representative is taken.  This implies that the
scattering amplitude will have a universal structure with a
combination-dependent coefficient; one degenerate combination will therefore
give zero regardless of whether or not the particular amplitude is
intrinsically non-vanishing.  Any other choice of linear combination
will reveal the true interaction.}

     The four-point functions (4.5) and (4.6) provide an illustration of
$W_3$ duality in the sense that they are zero because the underlying
three-point functions vanish given the particular $\varphi$ momenta of the
external states. For example, in (4.5) we could expand in the $s$ channel in
which case the intermediate states are from the $\leff_0=\ft{15}{16}$
sector. Since no such operators exist with $\beta=-\ft57 Q$ (see section 5
for a discussion on this point), there does not exist a three-point function
with $\V31{-\ft67 Q}{p_1}$ and $\V2{15/16}{-\ft37 Q}{p_2}$ as external
operators.

     The four-point function (4.7) is equivalent to (4.1$b$), after
interchanging the ordering of the operators.  Since the operators in (4.7)
and (4.1$b$) are different, this illustrates once more that interactions are
characterised by the effective spacetime structure of the physical
operators.  In addition, the comparison of (4.7) with (4.1$b$) illustrates
the crossing behaviour of $W_3$ scattering amplitudes, which in this
particular case amounts to interchanging $s$ with $t$, and $x$ with $(1-x)$.
Both these properties are also exhibited by (4.8), which is equivalent to
(4.1$e$) after interchanging $s$ with $u$, and $x$ with $1/x$.  The
computation of (4.8) is surprisingly simple, bearing in mind that the
operator $\V0{1/2}{\ft47Q}{p_3}$ given in ($A$.10) contains 82 terms.  This is
because making the replacement (2.14) selects just 27 terms, and of these
all but the 4 arising from those given explicitly in (3.5) do not
contribute.

     Let us now construct a five-point function.  This will provide further
indications that the observations we have made for three-point and four-point
functions are indeed general features of $W_3$-string theory.  In
particular, we shall exhibit the factorisation properties that are the
generalisation of duality in four-point functions.  Our example is built
from five physical operators that were given in [11], namely $\V2{15/16}{-\ft37
Q}{p_1}$, $\V2{15/16}{-\ft37 Q}{p_2}$, $\V2{1/2}{-\ft47 Q}{p_3}$,
$\V2{1/2}{-\ft47 Q}{p_4}$ and $\aV210{p_5}$ (which was inadvertently denoted
by $\V210{p_5}$ in [11]).  After some algebra, we find the
result
$$
\eqalign{
\fipf{\ft{15}{16}}{\ft{15}{16}}{\ft12}{\ft12}{1}&=
\int_0^1dx\, \int_0^1 dy\, x^{|p_1+p_2|^2/2 -2}
y^{|p_4+p_5|^2/2 -3/2}
(1-x)^{|p_2+p_3|^2/2 -31/16}\cr
&\qquad (1-y)^{|p_3+p_4|^2/2
-2}(1-xy)^{|p_2+p_4|^2/2 -31/16}(xy+x-2)\ ,\cr}
\eqno(4.9)
$$
where we have introduced the Koba-Nielsen variables $x$ and $y$, and $|p|^2
\equiv p\cdot (p+2a)$.  This example is particularly well suited to showing
both the similarities to, and the differences from, scattering amplitudes in
the ordinary bosonic string. For five tachyons in the ordinary string, one
would have
$$
\eqalign{
\int_0^1dx\, \int_0^1 & dy\, x^{(p_1+p_2)^2/2 -2}y^{(p_4+p_5)^2/2 -2}
(1-x)^{(p_2+p_3)^2/2 -2}\cr
& (1-y)^{(p_3+p_4)^2/2 -2}(1-xy)^{(p_2+p_4)^2/2 -2}\ .\cr}
\eqno(4.10)
$$
First, we note that (4.9) involves the background-charge vector $a_\mu$ in the
effective spacetime, which is of course absent in the usual 26-dimensional
bosonic string. both (4.9) and (4.10) display the property of
factorisation.  From (4.9) one can see for the $W_3$ string that states with
$\leff_0=1$, $\ft{15}{16}$ and $\ft12$ are exchanged in the various
intermediate channels, whereas the usual bosonic string amplitude (4.10)
shows that all the exchanged states have, of course, intercept 1.
Finally, we note the occurrence in (4.9) of the regular function $(xy+x-2)$.
This function, like the $(1-x+x^2)$ in (4.1$a$), and the $(x-2)$ in
(4.1$b$), comes from the functional integration over the spin-3 ghost system
$(\beta,\gamma)$ and the $\varphi$ field.  These functions are typical $W_3$
contributions to tree amplitudes.\footnote{$^*$}{\tenfoot These same
functions appear in the corresponding correlation functions of the Ising
model, if one makes the association described at the end of section 2.
However, as we discussed there, there is no one-to-one correspondence
between non-zero correlators in the $W_3$ string and the Ising model.  In
particular, one gets the function $\cos{\pi\over 8}\sqrt{1+\sqrt{1+x}}+
\sin{\pi\over 8}\sqrt{1-\sqrt{1-x}}$ from $\fpf\sigma\sigma\sigma\sigma$ in
the Ising model [20], whereas the function from
$\fpf{\ft{15}{16}}{\ft{15}{16}}{\ft{15}{16}}{\ft{15}{16}}$ in the $W_3$
string is 0, as follows from (4.3).  This is an indication that one cannot
simply view the $W_3$ string as a product of effective Virasoro strings with
the Ising model, since the latter gives a non-zero result (see [21]) for an
amplitude that is zero in the $W_3$ string.  Such a viewpoint does not
capture the essence of the $W_3$ symmetry, since it does not involve
functional integrations over all the quantum fields of the $W_3$ string.  In
particular, the selection rule (2.16) which comes from integration over the
$\varphi$ field, and is responsible for the vanishing of the four-point
function $\fpf{\ft{15}{16}}{\ft{15}{16}}{\ft{15}{16}}{\ft{15}{16}}$ in the
$W_3$ string, does not exist in such an approach.  In [21] this model of
effective Virasoro strings tensored with the Ising model emerged from
applying the ``group-theoretical method'' to the effective Virasoro strings.
Presumably if the method were applied instead to the $W_3$ string itself,
it would describe $W_3$-string scattering.}

     Not surprisingly, if we compute a five-point function $\fipf11111$ in
the $W_3$ string, we get the same result as (4.10), apart from the presence
of the background charge which means that $(p_i+p_j)^2$ is replaced by
$|p_i+p_j|^2$.  An example is provided by taking the operators
$\aV41{-\ft87Q}{p}$, $\V31{-\ft67Q}{p}$, and three $\V110{p}$ operators.

     We shall now explore the r\^ole of the discrete physical states in more
detail.  If one attempts to treat them on the same footing as physical
states with continuous momentum $p_\mu$, one finds that in $(N\ge4)$-point
functions they lead to divergent integrals.  This can be easily understood;
since the $p_\mu$ momentum of the discrete state is zero, it follows that
one can always use the duality or factorisation properties of the amplitude
to express it in terms of a sum over {\it on-shell} intermediate states at a
three-point vertex with the discrete state and another physical
state as external states, thereby giving a divergent result.  A simple
example is provided by looking at a four-point function in which one of the
operators is the level 4 discrete operator $\D00=1$, which corresponds to
the physical state $\ket{0}$.  One easily finds that the integral over $x$
is logarithmically divergent.  A less trivial example is provided by
considering the four-point function built from the operators $\aV210{p_1}$,
$\V2{15/16}{-\ft37Q}{p_2}$, $\aV3{15/16}{-\ft37Q}{p_3}$ and the discrete
operator $\D2{-\ft87Q}$. Following our procedure for calculating this
four-point function, we obtain the result
$$
\fpf1{\ft{15}{16}}{\ft{15}{16}}{D} = -120\int_0^1 dx\,\Big[3 + {2\over x(1-x)}
\Big]\ .\eqno(4.11)
$$
This has the same logarithmic divergence as one finds in the previous
example with the unit operator.

     These divergences can obviously be avoided by not performing the offending
integrations.  More precisely, in an $(N\ge4)$-point function in which $m$ of
the external physical states are discrete, no divergence will occur if only
$(N-3-m)$ replacements (2.14), and corresponding integrations, are
performed, rather than the usual $(N-3)$.  Let us consider the following
example, of a four-point function with the operators
$\V2{1/2}{-\ft47Q}{p_1}$, $\V2{1/2}{-\ft47Q}{p_2}$, $\aV210{p_3}$ and the
discrete operator $\D2{-\ft67Q}$.  Following the above prescription, there
will be no integration at all, and so {\it a priori} one might expect the
result to depend on the cross-ratio $x$.  Actually it does not, and the
result is simply the constant $\ft4{87}\sqrt{58}\,i$.  This is the result
that one would expect for a three-point function.  In fact this is precisely
what it is; what has happened is that the discrete state $\D2{-\ft67Q}$ has
normal ordered, by virtue of Wick's theorem, with another of the physical
operators to produce a new physical operator. Thus the net result is that
one is really just evaluating a three-point function of continuous-momentum
physical operators.

     This phenomenon generalises to any $N$-point function containing $m$
discrete operators.  Provided that they can normal order with some
continuous-momentum physical operators so as to produce new ones, then
Wick's theorem ensures that the $N$-point function reduces to an
$(N-m)$-point function of continuous-momentum physical operators.  If there
is no such normal ordering possible, then the result will be zero.  Indeed,
a set of discrete operators with $\varphi$ momenta $\beta_i$ has a
well-defined normal-ordered product with a continuous-momentum operator with
$\varphi$ momentum $\beta$ if and only if $\beta \sum_i \beta_i
+\sum_{i<j}\beta_i \beta_j$ is an integer.  Provided that the $m$ discrete
operators can be distributed over the original continuous-momentum operators
of the $N$-point function so that this condition is satisfied in each case,
then a well-defined $(N-m)$-point function exists; otherwise, it vanishes.
We have checked this in several examples.

     To end this section, we turn to a discussion of the special r\^ole of
the $\leff_0=\ft{15}{16}$ state in the physical spectrum of the $W_3$
string.  We have already seen that the selection rule (2.16) implies the
vanishing of the four-point function
$\fpf{\ft{15}{16}}{\ft{15}{16}}{\ft{15}{16}}{\ft{15}{16}}$ [11].  We shall now
argue that all $N$-point functions with three or more external physical
states from the $\leff_0=\ft{15}{16}$ sector are zero.

     We shall first demonstrate this result in the case when there are an odd
number of $\leff_0=\ft{15}{16}$ states in the $N$-point function.  It
follows from the fact that the $\varphi$ momentum of {\it any}
$\leff_0=\ft{15}{16}$ physical state is of the form $\beta=\ft{k}7Q$ where
$k$ is an odd integer, whereas for the $\leff_0=1$ and $\leff_0=\ft12$
sectors $k$ must be an even integer.  These properties can be proved from
the mass-shell condition implied by the fact that $L^{\rm tot}_0$
annihilates physical states, namely
$$
0=\ell-4-\ft12 \beta^2 -\beta Q +\leff_0\ ,\eqno(4.12)
$$
where $\ell$ is the level number.  Writing $\beta=\ft{k}7Q$, and recalling
that $Q^2=\ft{49}8$, this gives
$$
k^2+14k =16\leff_0+16\ell-64\ .\eqno(4.13)
$$
Now there is a general argument for {\it any} physical state that shows that
$k$ must certainly be an integer.  This can be seen by considering the case
when $\varphi$ is taken to be a timelike field, in which case it is
automatically periodic and hence its conjugate momentum is quantised [5,6] in
units of $\ft17 Q$.  (It can also be seen in all the physical
states that have been found in the $W_3$ string.)  Given that $k$ is an
integer, it is immediately clear that solutions of (4.13) when
$\leff_0=\ft{15}{16}$ require $k$ to be odd, whilst solutions when
$\leff_0=1$ or $\ft12$ require $k$ to be even.  Thus it is impossible to
satisfy the selection rule (2.16) if there is an odd number of
$\leff_0=\ft{15}{16}$ external states.

     For an even number of $\leff_0=\ft{15}{16}$ external states (or indeed
any number $\ge4$), we can make use of the factorisation property of the
$N$-point function.  This enables us to view the $N$-point function as a
``four-point function'' where three legs are external physical
$\leff_0=\ft{15}{16}$ states and the fourth is an intermediate state connecting
to the rest of the diagram.  Two distinct cases then arise.  If the
intermediate states in the fourth leg are in the $\leff_0=1$ or $\ft12$
sectors, then from the duality of the four-point function itself, it can be
seen to vanish because of the vanishing of its underlying three-point
functions.  If on the other hand the intermediate states of the fourth leg
are in the $\leff_0=\ft{15}{16}$ sector, then the four-point function
vanishes by virtue of (4.3).

     We have checked these properties in a number of examples.  More
specifically, we have verified explicitly that $N$-point functions such as
$\fipf{\ft{15}{16}}{\ft{15}{16}}{\ft{15}{16}}{\ft{15}{16}}{\ft12}$,
$\spf{\ft{15}{16}}{\ft{15}{16}}{\ft{15}{16}}{\ft{15}{16}}{\ft12}{\ft12}$,
and $\spf{\ft{15}{16}}{\ft{15}{16}}{\ft{15}{16}}{\ft{15}{16}}{\ft{15}{16}}
{\ft{15}{16}}$ all vanish.  This provides a non-trivial check, since there
are candidate amplitudes in which the selection rules (2.15) and (2.16) are
satisfied.

     The conclusion of the argument above is that any $N$-point function in
the $W_3$ string with three or more external $\leff_0=\ft{15}{16}$ states
vanishes, {\it i.e.}
$$
\big\langle \ft{15}{16}\ \ft{15}{16}\ \ft{15}{16}\ \cdots\big\rangle =0\ ,
\eqno(4.14)
$$
where $\cdots$ represents {\it any} set of physical states.  This shows
that indeed the $\leff_0=\ft{15}{16}$ sector plays a special r\^ole in the
$W_3$ string.

     In the $\leff_0=1$ and $\ft12$ sectors, on the other hand, no such
feature as (4.14) occurs.  Indeed it is possible to construct
non-vanishing scattering amplitudes with an arbitrarily large numbers of
physical operators from either of these sectors.  For $\leff_0=1$, the
$\ell=3$ prime operator $\V110{p}$ at ghost number $G=1$ can, with the
replacement (2.14), be inserted arbitrarily-many times in amplitudes without
disturbing the total $\varphi$ momentum or ghost number.  For
$\leff_0=\ft12$, one can in the same vein insert arbitrary numbers of pairs
of the physical operators $\V2{1/2}{-\ft47Q}{p}$ and $\V0{1/2}{\ft47Q}{p}$,
at levels $\ell=1$ and 8, and ghost numbers $G=2$ and 0.

     An interesting consequence of (4.14) is that at tree level
$\leff_0=\ft12$ states will not appear in intermediate channels if all the
external physical states lie in the $\leff_0=1$ and $\ft{15}{16}$ sectors.  In
loop amplitudes, on the other hand, states from all three sectors will run
around the loops.  This is illustrated by the fact that the one-loop partition
function of states with standard ghost structure (which all have $\leff_0=1$
or $\ft{15}{16}$) is not modular invariant [10].  This is to be expected,
since we know that the physical spectrum consists not only of the states with
standard ghost structure, but also of states with non-standard ghost
structure, which include in particular states in the $\leff_0=\ft12$ sector.

\bigskip
\bigskip
\noindent {\bf 5. More on the spectrum, and the no-ghost theorem}
\bigskip

\bigskip
\noindent{\it 5.1 The physical spectrum revisited}
\bigskip

    All the physical states with continuous momenta $p_\mu$ that we know of
display the remarkable property that they can be written as direct products
of a factor involving the ghost pairs $(b,c)$ and $(\beta,\gamma)$ together
with the frozen coordinate $\varphi$, and a factor involving only the $X^\mu$
coordinates, as in (3.7).  This means that all the known physical states with
continuous momentum admit an interpretation as effective spacetime physical
states. In other words, they are highest-weight states of $\teff$ given by
(2.5). It is worth noting that this is a highly non-trivial feature. For
example if we consider the $\ell=8$ prime state in, for simplicity, the
two-scalar case, there are 22 BRST-trivial structures at $G=-3$ which have to,
and indeed do, conspire to permit the removal of 134 terms that involve
excitations in the effective spacetime, thereby leaving 82 terms.  (Since, by
construction, these terms do not involve excitations in the effective
spacetime, their form, and number, is independent of the dimension of the
effective spacetime.  They are in fact precisely the terms given in
($A$.10).)  The occurrence of this phenomenon in numerous examples leads us
to conjecture that {\it all} physical states with continuous momentum
$p_\mu$ (prime states {\it and} quartet members) admit an effective
spacetime interpretation.

    The effective intercepts $\leff_0$ of the known continuous-momentum
physical states all take values in the set $\{1,\ \ft{15}{16},\ \ft12\}$.
The abundance of the known examples leads us to believe that this too is a
general feature of all the continuous-momentum physical states of the $W_3$
string.

    Given that the $\varphi$ momentum of physical states is quantised
in the form  $\ft{k}7 Q$, where $k$ is an integer, we can enumerate all the
possible level numbers and $\varphi$ momenta of all physical states which
admit an effective spacetime interpretation with $\leff_0$ lying in the set
$\{1,\ \ft{15}{16},\ \ft12\}$.  We shall present the argument for the case
where the effective spacetime state is a tachyon.  From the mass-shell
condition (4.13), it follows that
$$
k+7=\pm \sqrt{16\ell +16\leff_0-15}\ .\eqno(5.1)
$$
If we first consider the sector with $\leff_0=1$, we see that for $k$ to be
an integer it follows that we must have  $\ell=4p^2 +p$, with $p$ an
arbitrary integer.  Similar arguments can be applied to the other two
sectors, giving
$$
\eqalignno{
\leff_0=1&:\qquad \ell=4p^2+p,\qquad\qquad\,\, k=-7\pm (8p+1)\ ,&(5.2a)\cr
\leff_0=\ft{15}{16}&:\qquad \ell=p^2,\qquad\,\,\,\qquad\qquad k=-7\pm 4p\
,&(5.2b)\cr
\leff_0=\ft12&:\qquad \ell=4p^2+3p+1,\qquad k=-7\pm (8p+3)\ ,&(5.2c)\cr}
$$
where $p$ is an arbitrary integer in each case. It is instructive to
tabulate these results for the first few level numbers:

\bigskip
$$
\hbox{
\vbox{\tabskip=0pt \offinterlineskip
\def\tablerule{\noalign{\hrule}}
\halign to125pt{\strut#& \vrule#\tabskip=0em plus10em&
\hfil#\hfil& \vrule#&
\hfil#& \vrule#\tabskip=0pt\cr\tablerule
&&\multispan3 \hfil $\scriptstyle \leff_0=1$\hfil& \cr\tablerule
&&$\ell$&&${\scriptstyle k_-},\quad {\scriptstyle k_+}$&\cr \tablerule
&&${\scriptstyle 0}$&&$ {\scriptstyle -8},\quad {\scriptstyle -6}$\ &\cr
\tablerule
&&${\scriptstyle 3}$&&${\scriptstyle -14},\ \quad{\scriptstyle 0}$\ &\cr
\tablerule
&&${\scriptstyle 5}$&&${\scriptstyle -16},\ \quad{\scriptstyle 2}$\ &\cr
\tablerule
&&${\scriptstyle 14}$&&${\scriptstyle -22},\ \quad{\scriptstyle 8}$\ &\cr
\tablerule
&&${\scriptstyle 18}$&&${\scriptstyle -24},\quad{\scriptstyle 10}$\ &\cr
\tablerule
&&${\scriptstyle 33}$&&${\scriptstyle -30},\quad{\scriptstyle 16}$\ &\cr
\tablerule
&&${\scriptstyle 39}$&&${\scriptstyle -32},\quad {\scriptstyle 18}$\ &\cr
\tablerule \noalign{\bigskip}}}}\qquad
\hbox{
\vbox{\tabskip=0pt \offinterlineskip
\def\tablerule{\noalign{\hrule}}
\halign to125pt{\strut#& \vrule#\tabskip=0em plus10em&
\hfil#\hfil& \vrule#&
\hfil#& \vrule#\tabskip=0pt\cr\tablerule
&&\multispan3 \hfil ${\scriptstyle \leff_0=15/16}$\hfil& \cr\tablerule
&&$\ell$&&${\scriptstyle k_-},\quad {\scriptstyle k_+}$&\cr \tablerule
&&${\scriptstyle 0}$&&$ {\scriptstyle -7},\ \ {\scriptstyle -7}$\ &\cr
\tablerule
&&${\scriptstyle 1}$&&${\scriptstyle -11},\ \ {\scriptstyle -3}$\ &\cr
\tablerule
&&${\scriptstyle 4}$&&${\scriptstyle -15},\ \quad{\scriptstyle 1}$\ &\cr
\tablerule
&&${\scriptstyle 9}$&&${\scriptstyle -19},\ \quad{\scriptstyle 5}$\ &\cr
\tablerule
&&${\scriptstyle 16}$&&${\scriptstyle -23},\ \quad{\scriptstyle 9}$\ &\cr
\tablerule
&&${\scriptstyle 25}$&&${\scriptstyle -27},\quad{\scriptstyle 13}$\ &\cr
\tablerule
&&${\scriptstyle 36}$&&${\scriptstyle -31},\quad {\scriptstyle 17}$\ &\cr
\tablerule \noalign{\bigskip}}}}\qquad
\hbox{
\vbox{\tabskip=0pt \offinterlineskip
\def\tablerule{\noalign{\hrule }}
\halign to125pt{\strut#& \vrule#\tabskip=0em plus10em&
\hfil#\hfil& \vrule#&
\hfil#& \vrule#\tabskip=0pt\cr\tablerule
&&\multispan3 \hfil ${\scriptstyle \leff_0=1/2}$\hfil& \cr\tablerule
&&$\ell$&&${\scriptstyle k_-},\quad {\scriptstyle k_+}$&\cr \tablerule
&&${\scriptstyle 1}$&&${\scriptstyle -10},\ \ {\scriptstyle -4}$\ &\cr
\tablerule
&&${\scriptstyle 2}$&&${\scriptstyle -12},\ \ {\scriptstyle -2}$\ &\cr
\tablerule
&&${\scriptstyle 8}$&&${\scriptstyle -18},\ \quad{\scriptstyle 4}$\ &\cr
\tablerule
&&${\scriptstyle 11}$&&${\scriptstyle -20},\ \quad{\scriptstyle 6}$\ &\cr
\tablerule
&&${\scriptstyle 23}$&&${\scriptstyle -26},\quad{\scriptstyle 12}$\ &\cr
\tablerule
&&${\scriptstyle 28}$&&${\scriptstyle -28},\quad{\scriptstyle 14}$\ &\cr
\tablerule
&&${\scriptstyle 46}$&&${\scriptstyle -34},\quad{\scriptstyle 20}$\ &\cr
\tablerule \noalign{\bigskip}}}}
$$
\centerline{\it Table 3. Allowed levels and $\varphi$ momenta for effective
tachyons}
\bigskip

     The results presented in (5.2$a$--$c$) and the above table give the
level numbers $\ell$ of physical states in the $W_3$ string that are
tachyons from the effective spacetime point of view.  As we have discussed
earlier, one can always replace the tachyon by an excited effective
spacetime physical state with the same $\leff_0$ value.  If this effective
state has effective level number $\ell_{\rm eff}$, then the $\ell$ values
given above are shifted according to $\ell\rightarrow \ell +\ell_{\rm eff}$.

     The conditions given in (5.2$a$--$c$), and presented in Table 3,
represent {\it necessary} conditions, derived from the mass-shell
constraint, for the existence of continuous-momentum physical operators of the
kinds we are discussing.  There is no {\it a priori} reason to expect that
they must exist.  However comparison with the known physical operators (see
Table 1), shows that up to and including level 9, all the operators in Table 3
do actually occur, at the indicated $\leff_0$ values.  The $\varphi$ momenta
of all the physical operators in Table 1 correspond to the larger of the
allowed $k$ values, namely $\beta=\ft{k_+}7Q$; the $k_-$ value corresponds
to the conjugate physical operator in each case.  The operators with
standard ghost structure, at $\ell=0$, are an exception in that both $k_+$
and $k_-$ occur for both the operators and their conjugates.

     A similar discussion can be given for discrete states.  Although, as we
have explained in section 3, they do not in general admit an effective
spacetime interpretation (although the prime discrete states do) we may
still still use the mass-shell condition (4.13), with $\leff_0=0$ since they
have $p_\mu=0$.  Thus we find that they can occur at level numbers and
$\varphi$ momenta given by
$$
\ell=4p^2+p+1,\qquad k=-7\pm(8p+1)\ ,\eqno(5.3)
$$
where $p$ is an arbitrary integer.  The first few examples are:

\bigskip
$$
\hbox{
\vbox{\tabskip=0pt \offinterlineskip
\def\tablerule{\noalign{\hrule}}
\halign to200pt{\strut#& \vrule#\tabskip=0em plus10em&
\hfil#\hfil& \vrule#&
\hfil#& \vrule#\tabskip=0pt\cr\tablerule
&&$\ell$&&${\scriptstyle k_-},\quad {\scriptstyle k_+}$&\cr \tablerule
&&${\scriptstyle 1}$&&$ {\scriptstyle -8},\quad {\scriptstyle -6}$\ &\cr
\tablerule
&&${\scriptstyle 4}$&&${\scriptstyle -14},\ \quad{\scriptstyle 0}$\ &\cr
\tablerule
&&${\scriptstyle 6}$&&${\scriptstyle -16},\ \quad{\scriptstyle 2}$\ &\cr
\tablerule
&&${\scriptstyle 15}$&&${\scriptstyle -22},\ \quad{\scriptstyle 8}$\ &\cr
\tablerule
&&${\scriptstyle 19}$&&${\scriptstyle -24},\quad{\scriptstyle 10}$\ &\cr
\tablerule
&&${\scriptstyle 34}$&&${\scriptstyle -30},\quad{\scriptstyle 16}$\ &\cr
\tablerule
&&${\scriptstyle 40}$&&${\scriptstyle -32},\quad {\scriptstyle 18}$\ &\cr
\tablerule \noalign{\bigskip}}}}
$$
\centerline{\it Table 4. Allowed levels and $\varphi$ momenta}
\centerline{\it $\quad$ for discrete operators}
\bigskip

      The level $\ell=1$, 4 and 6 discrete operators in the table are the
ones we discussed in section 3. In fact, as we explained there, we can take
normal ordered products of the discrete operators to create new ones. From
equation (3.10) we see that discrete operators in Table 4 at
$\ell\ge 15$ with $\varphi$ momentum $\ft{k_+}7Q$ can be obtained from
normal ordered powers of the $\ell=6$ operator. Similarly by normal ordering
discrete operators with continuous-momentum physical operators, new such
operators are created. In a certain sense, a discrete operator can thus be
viewed as a ``screening charge'' in the $\varphi$ direction times the
identity operator in the effective spacetime directions.   The discrete
operator $\D0{\ft27Q}$ at $\ell=6$ and $G=0$ given in ($A$.14) provides a
nice illustration.  Since it has $\varphi$ momentum $\beta=\ft27Q$, it could
in principle enable one to step up through the $k_+$ values of the three
sectors in Table 3.  Since physical operators with non-negative $\varphi$
momentum will certainly have a non-vanishing normal-ordered product with
appropriate powers of $\D0{\ft27Q}$, it may be that one should regard the
first level number in each sector at which $k_+$ is non-negative as a
foundation from which higher levels in that sector can be obtained.  As an
example, we have explicitly checked that normal ordering $\D0{\ft27Q}$ given
in ($A$.14) with the $\ell=3$ prime operator ($A$.7) of the $\leff_0=1$
sector gives the $\ell=5$ prime operator ($A$.9).

\bigskip
\noindent{\it 5.2 Unitarity of the $W_3$ string}
\bigskip

     Let us now consider the question of the unitarity of the $W_3$ string.
For physical states of standard ghost structure, unitarity was first proven
in [6].  The argument consists of observing that since such states all admit an
effective spacetime interpretation (see (2.6) and (2.10)), proving
unitarity for these states amounts to proving unitarity for an effective
Virasoro string theory with central charge $c=\ft{51}2$ and intercepts $a=1$
and $a=\ft{15}{16}$.  A standard result from ordinary string theory is that
the unitarity bounds derived from level 1 and level 2 states are sufficient
to ensure unitarity at all excited levels.  As discussed, for example, in
[6], the unitarity of level states requires $a\le 1$, whilst level 2
unitarity requires
$$
a\le {37-c-\sqrt{(c-1)(c-25)}\over 16}\qquad {\rm or}\qquad
a\ge {37-c+\sqrt{(c-1)(c-25)}\over 16}\ .\eqno(5.4)
$$
For the case of $c=\ft{51}2$, these bounds imply that the effective
spacetime intercept must satisfy
$$
a\le \ft12\qquad {\rm or}\qquad \ft{15}{16}\le a \le 1\ .\eqno(5.5)
$$
Thus the physical states in the $W_3$ string with standard ghost structure
precisely saturate the limits of the second interval.  This means that they
are unitary.  (In fact because they saturate the limits, it means that there
are null states, characteristic of spacetime gauge symmetries, as well as
positive-norm states.)

     This reasoning extends to {\it all} physical states in the $W_3$ string
that admit an effective spacetime interpretation.  All the known examples of
continuous-momentum physical states do indeed admit such
an interpretation, and furthermore, their effective intercept values lie in
the set $\{1,\ \ft{15}{16},\ \ft12\}$.  Thus all the known
continuous-momentum physical states in the $W_3$ string satisfy the
unitarity conditions (5.5).  In fact they all saturate one or another of the
limits in (5.5).

     If, as we have conjectured in this paper, all continuous-momentum
physical states of the $W_3$ string admit an effective spacetime
interpretation, with $\leff_0$ values lying in the set $\{1,\ \ft{15}{16},\
\ft12\}$, it follows that the argument above constitutes a proof of
unitarity of the $W_3$ string.

     Discrete states do not upset this conclusion.  This is because, unlike
the continuous-momentum physical states whose unitarity we have already
addressed, a discrete state is a scalar under the effective-spacetime
Lorentz group rather than a tensor, and so the issue of negative-norm states
does not arise.

     Another indication of the unitarity of the $W_3$ string can be seen by
looking at the residues of the poles in intermediate channels of $(N\ge
4)$-point functions.  Let us consider first the example of the four-point
function $\fpf1111$ given in (4.4).  (This is of the same form as the
four-point tachyon amplitude in the bosonic string.)  The integral can be
written as
$$
\fpf1111 \sim
{\Gamma(-s/2-1)\Gamma(-t/2-1)\over \Gamma(-s/2-t/2-2)}\ .\eqno(5.6)
$$
Using the standard expression for the behaviour of the $\Gamma$ function
near a pole, $\Gamma(-n+\epsilon)=(-1)^n/n!\, \epsilon^{-1}$, and
expanding in the $t$-channel poles, $t/2+1=n-\epsilon$, we find for the
leading-order in $s$:
$$
\fpf1111 \sim
\sum_{n=0}^\infty {1\over n!}\, {(s/2)^n\over n-1-t/2} \ .
\eqno(5.7)
$$
The fact that residues at all the poles have the same sign implies that the
propagators of all the intermediate states in the sum have the same sign,
and correspondingly they do not have negative norm.  Similar arguments can
be applied to the other channels, and to all the other $N$-point functions
of the $W_3$ string.  For example, expanding the
$\fpf{\ft12}{\ft12}{\ft12}{\ft12}$ result given in (4.1$a$) in the $t$
channel gives
$$
\fpf{\ft12}{\ft12}{\ft12}{\ft12} \sim
\sum_{n=1}^\infty {1\over n!}\, {(s/2)^n \over n-1-t/2}
\eqno(5.8)
$$
to leading order in $s$.

\bigskip
\bigskip
\noindent {\bf 6. Mixing the $\varphi$ and ghost fields}
\bigskip

     The fact that all the physical states with continuous momentum admit an
effective spacetime interpretation suggests that there may be a more
appropriate description of the $W_3$ string, where this feature is more
manifest.  Motivated by this, we start by noting that we can write the BRST
operator $Q_B$ of (2.1) as
$$
Q_B=\oint dz\Big[ \widetilde c\, \teff + {\rm more}\Big]\ ,\eqno(6.1)
$$
where all the dependence on the spacetime coordinates $X^\mu$ is contained
in the first term, and
$$
\widetilde c\equiv c+\ft7{174}\sqrt{58}\,i\, \del\gamma -\ft8{261} b\,
\del\gamma\, \gamma -\ft4{87}\sqrt{29}\,i\, \del\varphi\, \gamma \ .\eqno(6.2)
$$
This suggests that simplifications might occur if we were to treat
$\widetilde c$ rather than $c$ as the ghost for the spin-2 symmetry. We thus
look for a corresponding set of redefinitions for $b$, $\gamma$, $\beta$ and
$\varphi$ such that the new fields have the same set of operator-product
expansions as the original ones.  This leads uniquely to
$$
\eqalign{
\widetilde c&\equiv c+\ft7{174}\sqrt{58}\,i\, \del\gamma -\ft8{261} b\,
\del\gamma\, \gamma -\ft4{87}\sqrt{29}\,i\, \del\varphi\, \gamma\cr
\widetilde b &\equiv b\ ,\cr
\widetilde \gamma &\equiv \gamma\ ,\cr
\widetilde \beta &\equiv \beta + \ft7{174}\sqrt{58}\,i\, \del b
-\ft8{261}\del b\, b\, \gamma +\ft4{87}\sqrt{29}\,i\, \del\varphi\, b\ ,\cr
\widetilde\varphi &\equiv \varphi-\ft4{87}\sqrt{29}\,i\, b\,\gamma\ .\cr}
\eqno(6.3)
$$
These redefinitions may be inverted, to give
$$
\eqalign{
c &\equiv \widetilde c-\ft7{174}\sqrt{58}\,i\, \del\widetilde \gamma
-\ft8{261} \widetilde b\, \del\widetilde \gamma\, \widetilde \gamma
+\ft4{87}\sqrt{29}\,i\, \del\widetilde \varphi\, \widetilde \gamma \ ,\cr
b &\equiv \widetilde b\ ,\cr
\gamma &\equiv \widetilde \gamma\ ,\cr
\beta &\equiv \widetilde \beta - \ft7{174}\sqrt{58}\,i\, \del \widetilde b
-\ft8{261}\del \widetilde b\, \widetilde b\, \widetilde \gamma
-\ft4{87}\sqrt{29}\,i\, \del\widetilde \varphi\, \widetilde b\ ,\cr
\varphi &\equiv \widetilde \varphi+\ft4{87}\sqrt{29}\,i\, \widetilde b\,
\widetilde \gamma\ .\cr}\eqno(6.4)
$$
Since they preserve the operator products, the transformations
(6.3) are in some sense canonical.

     It is now a straightforward matter to use (6.4) to rewrite the BRST
operator, the ghost boosters $a_\varphi$ and $a_{X^\mu}^\dum$, and the
physical operators in terms of the new fields.  We shall first consider the
physical operators.  Remarkably, the higher-level physical operators
become greatly simplified.  The results for the physical
operators with continuous momentum presented in the appendix become:

\noindent$\bullet\ \underline{\ell=0,\ G=3}$:
$$
\eqalignno{
\V{3}{1}{-\ft87 Q}{p} &=\widetilde c\, \del \widetilde \gamma\, \widetilde
\gamma\, e^{-\ft87 Q \tilde\varphi}e^{i p\cdot X}\ ,&(6.5)\cr
\V{3}{1}{-\ft67 Q}{p} &=\widetilde c\, \del \widetilde \gamma\,\widetilde
\gamma \, e^{-\ft67 Q \tilde\varphi}e^{i p\cdot X}\ ,&(6.6)\cr
\V{3}{15/16}{-Q}{p} &=\widetilde c\, \del\widetilde \gamma\,\widetilde
\gamma \, e^{-Q \tilde\varphi}e^{i p\cdot X}\ .&(6.7) \cr}
$$

\noindent$\bullet\ \underline{\ell=1,\ G=2}$:
$$
\eqalignno{
\V{2}{15/16}{-\ft37 Q}{p} &=\widetilde c\,\widetilde \gamma e^{-\ft37 Q
\tilde\varphi} e^{ip\cdot X}\ ,&(6.8)\cr
\V{2}{1/2}{-\ft47 Q}{p} &=\widetilde c\,\widetilde\gamma e^{-\ft47 Q
\tilde\varphi} e^{ip\cdot X}\ .&(6.9)\cr}
$$

\noindent$\bullet\ \underline{\ell=2,\ G=2}$:
$$
\V{2}{1/2}{-\ft27 Q}{p} = \widetilde c\Big(\sqrt{2}\,\del\widetilde \varphi\,
\widetilde \gamma -\ft32\del\widetilde\gamma\Big)
e^{-\ft27 Q \tilde\varphi} e^{i p\cdot X}\ .\eqno(6.10)
$$

\noindent$\bullet\ \underline{\ell=3,\ G=1}$:
$$
\V{1}{1}{0}{p}=
\widetilde c\, e^{ip\cdot X}\ .
\eqno(6.11)
$$

\noindent$\bullet\ \underline{\ell=4,\ G=1}$:
$$
\V{1}{15/16}{\ft17Q}{p} =-\ft3{16}\sqrt{29}\,i\,\widetilde c\,
\Big(3\sqrt{2}\,
\widetilde\beta\, \widetilde\gamma  +4\, \del \widetilde\varphi
\Big) e^{\ft17 Q \tilde\varphi} e^{ip\cdot X}\ .\eqno(6.12)
$$

\noindent$\bullet\ \underline{\ell=5,\ G=1}$:
$$
\V{1}{1}{\ft27Q}{p} = \ft1{20}\sqrt{29}\,i\, \widetilde c\,
\Big( 6\sqrt{2}\, \del\widetilde \beta\,\widetilde \gamma  -3\sqrt{2}\,
\widetilde \beta\,\del\widetilde \gamma+
12 \del\widetilde \varphi\, \widetilde \beta\,
\widetilde \gamma+ 4\sqrt{2}\, \del\widetilde
\varphi\,\del\widetilde \varphi\,
+ 2 \del^2\widetilde \varphi\,  \Big)e^{\ft27 Q\tilde \varphi}
e^{ip\cdot X}\ .\eqno(6.13)
$$

\noindent$\bullet\ \underline{\ell=8,\ G=0}$:
$$
\V{0}{1/2}{\ft47Q}{p}= \widetilde c\,\Big( \del \widetilde \varphi\, \del
 \widetilde \beta + 3\sqrt{2}
\,\del \widetilde \beta\, \widetilde \beta\, \widetilde \gamma -
3\, \del^2 \widetilde \varphi \, \widetilde \beta -\ft34\sqrt{2}
\,\del^2 \widetilde \beta \Big)e^{\ft47Q \tilde \varphi}
e^{ip\cdot X}\ .\eqno(6.14)
$$

\noindent$\bullet\ \underline{\ell=9,\ G=0}$:
$$
\eqalign{
\V0{15/16}{\ft57Q}{p}=\widetilde c\, \Big(& \del\widetilde\varphi\,
\del\widetilde\varphi \,\del\widetilde\beta -\ft38\sqrt{2}\,\del
\widetilde\varphi\,\del^2\widetilde\beta -2\sqrt{2}\,
\del^2\widetilde\varphi\, \del\widetilde\beta -3\, \del^2\widetilde\varphi
\,\del\widetilde\varphi\,\widetilde\beta -\ft38\sqrt{2}\,
\del^3\widetilde\varphi\,\widetilde\beta \cr
&+\ft{15}4\sqrt{2}\, \del\widetilde\varphi\, \del\widetilde\beta\,
\widetilde\beta\, \widetilde\gamma -3\, \del\widetilde\beta\,
\widetilde\beta\, \del\widetilde\gamma + \ft{27}{16} \del^2\widetilde\beta
\, \widetilde\beta \, \widetilde\gamma -\ft12 \del^3\widetilde\beta \Big)
e^{\ft57 Q \tilde\varphi} e^{ip\cdot X}\ .}\eqno(6.15)
$$
The $\ell=1$ and $\ell=6$ discrete operators given in the appendix become

\noindent$\bullet\ \underline{\ell=1,\ G=2}$:
$$
\eqalignno{
\D{2}{-\ft67Q}&= \Big(\widetilde c\, \widetilde \gamma
+\ft2{87}\sqrt{58}\,i\, \del\widetilde \gamma\, \widetilde \gamma
\Big) e^{-\ft67 Q\tilde \varphi}\ ,&(6.16)\cr
\D{2}{-\ft87Q}&= \Big(\widetilde c\, \widetilde \gamma
+\ft5{87}\sqrt{58}\,i\, \del\widetilde \gamma\, \widetilde \gamma
\Big) e^{-\ft87 Q\tilde\varphi}\ .&(6.17)\cr}
$$

\noindent$\bullet\ \underline{\ell=6,\ G=0}$:
$$
\D0{\ft27Q}=
\Big[\ft{261}{10} \widetilde c\, \widetilde\beta
-\ft15\sqrt{29}\,i\,\Big(6\sqrt{2}\, \del\widetilde \beta\,\widetilde \gamma
 -3\sqrt{2}\, \widetilde \beta\,\del\widetilde \gamma+ 12 \del\widetilde
\varphi\, \widetilde \beta\, \widetilde \gamma+ 4\sqrt{2}\, \del\widetilde
\varphi\,\del\widetilde \varphi\, + 2 \del^2\widetilde \varphi\,\Big)
\Big]e^{\ft27 Q\tilde \varphi}\ .
\eqno(6.18)
$$

    Several comments are now in order. In all the above expressions, the
implicit normal ordering is with respect to the tilded fields.  Since the
$(b\,\gamma)$ term in the redefinition of $\varphi$ is nilpotent, the
exponential terms of the form $e^{\lambda \tilde \varphi}$ reduce to
$$
\Big(1-\ft4{87}\sqrt{29}\,i\, \lambda\, b\,\gamma\Big)e^{\lambda\varphi}
\eqno(6.19)
$$
in terms of the original fields.

     It is instructive to compare the physical operators as given above with
their expressions in terms of the original fields given in the appendix.  In
particular, the comparisons between the expressions (6.13) and ($A$.9) for the
$\ell=5$ operator, where 32 terms reduce to 5, and between the expressions
(6.14) and ($A$.10) for the $\ell=8$ operator, where 82 terms reduce to 4, are
quite striking.  The level $\ell=9$ operator given in (6.15) does not appear
in the appendix.  In fact we have obtained it only using the new formalism
of this section.  In the original formalism it involves 414 terms, thus
providing another example of the power of the new formalism.

     The calculation of scattering amplitudes in the $W_3$ string can be
performed equivalently in terms of the new quantum fields given by
(6.3).  Since the transformed fields satisfy the same operator-product
expansions as the original ones, the Jacobian of the transformation is
simply the identity, and one can calculate with the tilded fields in
exactly the same way as we did with the untilded ones.  For example,
(2.15) is identical in its tilded form.  Interestingly, we see that the
terms in (6.12), (6.13) and (6.14) are of precisely the same form as those in
(3.1), (3.3) and (3.5), which were the only terms in ($A$.8), ($A$.9) and
($A$.10) that contributed in scattering amplitudes.  This explains the high
degree of redundancy in the representation of physical states in terms of
the original fields, where many of the terms in a physical state seemed to
be irrelevant in interactions.  The new fields $\widetilde c$, $\widetilde
b$, $\widetilde\gamma$, $\widetilde\beta$ and $\widetilde \varphi$ thus
provide what one might call a ``minimal'' formalism in which this
redundancy is removed.

     All the above physical operators with continuous momentum $p_\mu$ have
the form
$$
U(\widetilde\beta,\widetilde\gamma,\widetilde\varphi)\; \widetilde c\,
e^{ip\cdot X}\ .\eqno(6.20)
$$
The $\widetilde c\, e^{ip\cdot X}$ factor is reminiscent of an effective
string theory. We conjecture that {\it all} prime physical operators with
continuous momentum $p_\mu$ have the form (6.20) (or with $\del\,\widetilde
c\,\widetilde c$ in the case of the conjugate physical operator).   The
discrete operators, on the other hand, do not factorise in this way.

     The ghost boosters $a_\varphi$ and $a_{X^\mu}^\dum$ take on simpler
forms in terms of the new fields:
$$
\eqalign{
a_{\tilde\varphi}&\equiv [Q_B,\widetilde\varphi] = \widetilde c\,\del
\widetilde\varphi -Q\, \del\,\widetilde c\cr
&\qquad\qquad\quad +\ft1{261}\sqrt{29}\,i\,\Big( 19\, \del^2\widetilde\gamma
+36\,\widetilde\beta\, \del\widetilde\gamma\,\widetilde\gamma
+24\,\del\widetilde\varphi\,\del\widetilde\varphi\,\widetilde\gamma
-24Q\, \del\widetilde\varphi\, \del \widetilde\gamma\Big)\ ,\cr
a_{X^\mu}^\dum&\equiv [Q_B,X^\mu]
=\widetilde c\, \del X^\mu - i\, a^\mu \del\,\widetilde c\ .\cr}\eqno(6.21)
$$
Note that $a_{\tilde\varphi}$ is equivalent to $a_\varphi$ modulo a
BRST-trivial term, since $[Q_B,b\,\gamma]$ is BRST trivial.  It is clear
from (6.21) why the $a_{X^\mu}^\dum$ ghost boosters generate just one physical
operator at ghost number $(G+1)$ by acting on a prime operator at ghost number
$G$, as we described in section 3:  for an operator of the form (6.20), we
get
$$
a_{X^\mu}^\dum \,
U(\widetilde\beta,\widetilde\gamma,\widetilde\varphi)\; \widetilde c\,
e^{ip\cdot X} = i\, (-1)^G (p^\mu +a^\mu)\,
U(\widetilde\beta,\widetilde\gamma,\widetilde\varphi)\; \del\, \widetilde
c\,\widetilde c\, e^{ip\cdot X}\ .\eqno(6.22)
$$
{}From the form of the ghost boosters (6.21), it becomes clear that the entire
quartet generated from the physical operator (6.20) will admit an effective
spacetime interpretation.  Since, on the other hand, the discrete operators
cannot be written in the form (6.20),  $a_{X^\mu}^\dum$ generates $d$
independent discrete operators rather than just one. Correspondingly, these
boosted operators do not all admit an effective spacetime interpretation.

     The introduction of the tilded fields makes manifest an observation
that we made previously in section 2.  There, we observed that in
ordinary string theory one would get zero in $(N\ge4)$-point scattering
amplitudes unless the replacement (2.14) is made for $(N-3)$ of the
operators, since each operator involves a $c$ ghost, but that it was {\it a
priori} not so clear that the same would be true in the $W_3$ string.
However, as we have now seen by using the tilded fields, all known physical
operators with continuous momentum involve a factor of $\widetilde c$, and
thus the replacement (2.14) is necessary for the same reason as in ordinary
string theory.  Since discrete operators do not have this structure, the
same reasoning does not apply, which is consistent with our findings in
section 4 where the replacement (2.14) had to be omitted in order to get
finite and non-zero results in amplitudes involving them.

     Another feature of the $W_3$ string that becomes manifest in the new
formalism is that tree-level $N$-point amplitudes where all the external
states have $\leff_0=1$ are identical to those of the ordinary bosonic
string.  As an example, for effective tachyons we can take two $\ell=0$
operators $\aV41{-\ft87Q}{p}$ and $\V31{-\ft67Q}{p}$, and $(N-2)$ operators
$\V110{p}$ given in (6.11).  As usual, one could replace the tachyons by
excited effective spacetime states if desired.

     Finally, we give the form of the BRST operator in terms of the new
fields:
$$
\eqalign{
Q_B=Q_B^{\rm eff} +&\oint dz\, \Big[ \widetilde c\, T_{\tilde\varphi} -
3\,\widetilde c\,\widetilde\beta\,\del\widetilde\gamma -2\, \widetilde c\,
\del\widetilde\beta\, \widetilde\gamma\cr
-\ft4{261} & \sqrt{29}\,i\, \widetilde \gamma\,
\Big(2(\del\widetilde\varphi)^3 +6Q\,\del^2\widetilde\varphi\,\del
\widetilde\varphi +\ft{19}4\,\del^3\widetilde\varphi + 9\,\del\widetilde
\varphi\, \widetilde\beta\,\del\widetilde\gamma + 3Q\, \del\widetilde\beta\,
\del\widetilde\gamma\Big)\Big]\ ,\cr}\eqno(6.23)
$$
where
$$
Q_B^{\rm eff}\equiv \oint dz\, \widetilde c\Big(\teff +\ft12 T_{\tilde
c,\tilde b}\Big)\ ,\eqno(6.24)
$$
and
$$
\eqalignno{
T_{\tilde\varphi}&\equiv -\ft12(\del\widetilde\varphi)^2 -Q\, \del^2
\widetilde\varphi\ ,&(6.25)\cr
T_{\tilde c,\tilde b}&\equiv -2\,\widetilde b\,\del\, \widetilde c -
\del\widetilde b\, \widetilde c\ .&(6.26)\cr}
$$

     Note that although $Q_B^{\rm eff}$ is not nilpotent, since $\teff$ has
central charge $\ft{51}2$ and $T_{\tilde c,\tilde b}$ has central
charge $-26$, it does have the property that it is nilpotent on the subspace
spanned by $\widetilde c$, $\widetilde \gamma$, $\widetilde \beta$,
$\widetilde \varphi$ and $X^\mu$, but not on $\widetilde b$, in the sense that
$$
[Q_B^{\rm eff},\{Q_B^{\rm eff},\widetilde b\, \}]=-\ft1{24} \del^3
\widetilde c\ .\eqno(6.27)
$$

     Although $Q_B^{\rm eff}$ is not itself nilpotent, by adding all the
other terms involving $\widetilde c$ in (6.23) we obtain a nilpotent
operator $Q_0$. In fact writing $Q_B=Q_0+Q_1$, with
$$
\eqalign{
Q_0&=Q_B^{\rm eff}+\oint dz\,\Big( \widetilde c\, T_{\tilde\varphi}-
3\, \widetilde c\,\widetilde\beta\,\del\widetilde\gamma -
2\,\widetilde c\,\del\widetilde\beta\,\widetilde\gamma\Big)\cr
&=\oint dz\, \widetilde c \Big( T^{\rm eff}+T_{\tilde\varphi}+
T_{\tilde\gamma,\tilde\beta}+\ft12 \,
T_{\tilde c,\tilde b}\Big)\ ,\cr
\cr
Q_1&=-\ft4{261} \sqrt{29}\,i\, \oint dz\,
\widetilde \gamma\,
\Big(2(\del\widetilde\varphi)^3 +6Q\,\del^2\widetilde\varphi\,\del
\widetilde\varphi +\ft{19}4\,\del^3\widetilde\varphi + 9\,\del\widetilde
\varphi\, \widetilde\beta\,\del\widetilde\gamma + 3Q\, \del\widetilde\beta\,
\del\widetilde\gamma\Big)\ ,\cr}\eqno(6.28)
$$
where
$T_{\tilde\gamma,\tilde\beta}=-3\,\widetilde\beta\,\del\widetilde\gamma-
2\, \del\widetilde\beta\,\widetilde\gamma$, the BRST charge separates
into two independently nilpotent parts. The nilpotency of $Q_0$ is easy to see
because the total central charge of $ T^{\rm eff}+T_{\tilde\varphi}+
T_{\tilde\gamma,\tilde\beta}$ is 26, and the nilpotency of $Q_1$ follows
from that of $Q_B$ and $Q_0$. The splitting is very natural if we view it
from another angle: $Q_0$ and $Q_1$ have different $(\widetilde c,
\widetilde b)$ and $(\widetilde\gamma,\widetilde\beta)$ ghost numbers. $Q_0$
has
$(\widetilde c,\widetilde b)$ ghost number 1 and $(\widetilde\gamma,
\widetilde\beta)$ ghost number 0, while $Q_1$ has $(\widetilde c ,\widetilde
b)$ ghost number 0 and $(\widetilde\gamma,\widetilde\beta)$ ghost number 1.
Thus we have
$$
\eqalign{
Q_B&=Q_0+Q_1\ ,\cr
Q_0^2&=Q_1^2=\big\{Q_0,\ Q_1\big\}=0\ .\cr}\eqno(6.29)
$$
This grading of the BRST operator provides an elegant starting point for
studying its cohomology.

\bigskip
\bigskip
\noindent {\bf 7. Conclusions}
\bigskip

     In this paper we have investigated in detail the structure of the
spectrum and the interactions of the open $W_3$ string.  We have seen that
the physical spectrum displays a surprising variety of states. As well as
the physical states of standard ghost structure, at ghost number $G=0$,
which have been known for a long time, it has been found that it also
contains prime states at ghost numbers $G=-1$, $G=-2$ and $G=-3$, and
discrete prime states at $G=-1$ and $G=-3$.  The prime states at $G=-2$ were
found for the first time in this paper.  We have reason to suspect that this
is not the end of the story; it seems likely that physical states at more
negative ghost numbers can exist too, both with continuous and with discrete
momenta.  We have also clarified the structure of the multiplets generated
by the action of the ghost boosters $a_\varphi$ and $a_{X^\mu}^\dum$ on the
prime states. For continuous-momentum prime states they generate a quartet
of physical states, at ghost numbers $\{G,\ G+1,\ G+2\}$ with multiplicities
$\{1,\ 2,\ 1\}$. For discrete prime states they generate a $(4d)$-plet of
discrete states at ghost numbers $\{G,\ G+1,\ G+2,\ G+3\}$ with
multiplicities  $\{1,\ d+1,\ 2d-1,\ d-1\}$.

     All of the $N$-point functions that we have calculated in [11] and in
this paper have involved physical states that are tachyons from the
effective-spacetime point of view.  As we have pointed out already, these
effective tachyons can be replaced by excited effective physical states with
the same $\leff_0$ values.  (It is for states such as these that our
unitarity discussion in section 5 applied.)  The scattering amplitudes that
we have presented can easily be modified to cover the case where such
physical states with excitations in the effective spacetime are involved.
This is because the excited physical states still have the form (3.7), and
so the contribution from the $\varphi$ and ghost fields will be the same as
in the purely tachyonic case.  Thus the standard techniques developed for
calculating tree amplitudes for excited states in the bosonic string can be
carried over to the $W_3$ string.

     All the known physical states with continuous momentum $p_\mu$ admit an
effective spacetime interpretation as given in (3.7), and have effective
intercepts $\leff_0=1$, $\ft{15}{16}$ or $\ft12$.  We have conjectured that
these features will hold for all physical states in the $W_3$ string with
continuous $p_\mu$ momentum.

     An important consequence of the fact that physical states with
continuous momentum $p_\mu$ admit such an effective spacetime interpretation
is that scattering amplitudes factorise into two parts.  The first part
involves only the effective spacetime states and the second involves only the
ghost fields and the $\varphi$ coordinate.  The result of the second part,
and therefore the entire amplitude, is universal in the sense that for a
given set of $\leff_0$ values for the external states, the result from the
ghost and $\varphi$ fields is always the same, up to a constant coefficient.
This coefficient is zero if any selection rule is violated.  We have
encountered several selection rules in this paper.  One is that the total
ghost structure must be of the form (2.15); another is the
$\varphi$-momentum conservation law (2.16); and another is that an $(N\ge
4)$-point function will vanish if any of its underlying three-point
functions vanishes.

     It is very striking that the frozen coordinate $\varphi$ is more akin
to the ghost fields than to the effective coordinates $X^\mu$.  In fact the
momentum conservation law (2.16) is rather similar to the condition that the
total ghost structure must be as in (2.15) to have a non-vanishing inner
product. This similarity becomes particularly evident if one bosonises
the ghosts.  In fact in section 6, we have shown that by performing a
redefinition of the quantum fields in which the ghosts and the $\varphi$
field are mixed, great simplifications occur.  In our previous paper on the
interacting $W_3$ string, we argued that the existence of physical states
involving excitations of the ghost fields was suggestive of the fact that
it may be inappropriate to look for a classical correspondence principle for
the $W_3$ string [11]. The results of this paper, and particularly of
section 6, reinforce this suggestion.  They also strengthen our belief that
a complete understanding of $W$-geometry can only be achieved at the quantum
level, when the ghosts are included.

     In this paper we have been concerned only with tree-level amplitudes.
The evaluation of loop amplitudes remains an interesting open question. We
expect that some new features should arise. For example, one of the results
in this paper is that the tree level amplitudes involving only $\leff_0=1$
states are the same as in the ordinary bosonic string. At loop level,
however, states with $\leff_0=\ft{15}{16}$ and $\ft12$  will presumably also
run around the loops, giving results that may differ from those of string
theory. Since the $\leff_0=1$ sector of the $W_3$ string includes massless
states, we may thus expect that the low energy limit of the $W_3$ string
might differ from that of the usual string.

\bigskip\bigskip
\centerline{\bf ACKNOWLEDGMENTS}
\bigskip

     We are grateful to John Dixon, Mike Duff and Peter West for
discussions.  Stany Schrans is indebted to the Center for Theoretical
Physics at Texas A\&M University for hospitality and support.  We have made
extensive use of the Mathematica package OPEdefs [22] written by Kris
Thielemans for the computations in sections 3 and 6.

\vskip1truein
\noindent{\bf NOTE ADDED}
\bigskip

     After this preprint was circulated, a paper by Freeman and West on
related  topics appeared [23].  They assert that the procedure we
introduced in [11] for calculating  $W_3$ string scattering amplitudes is
incorrect, and that it gives results  that violate unitarity and the
assumptions of S-matrix theory.  This is not  the case, and in fact the
authors of [23] explicitly use the procedure of [11] for  computing $W_3$
scattering amplitudes, including our use of ghost boosters and the
replacement (2.14) in order to achieve the correct ghost structures for
non-vanishing correlation functions.

     The authors of [23] introduce a screening charge $S=\oint dz\, U(z)$
in  order to obtain further scattering amplitudes beyond those that we
have  computed.  It is easy to see that this screening charge is related
by a descent equation to our $\ell=6$ discrete state $D^0[\ft27Q]$ given
in (6.18). In fact, in the new formalism introduced in section 6 of this
paper, their screening current $U$, after removing a total derivative, is
simply
given by $U=\widetilde \beta\, e^{\ft27 Q\tilde \varphi}$, and satisfies
$\{Q_B,U\}=\ft{10}{261} \del D^0[\ft27Q]$.  Conversely, we see that $U$ can
be obtained from $D^0[\ft27Q]$ by making the replacment (2.14).  In fact the
screening charge $S$ given in [23] is nothing but a special case of our
standard replacement of a spin 0 physical state by a spin 1 current that can
be integrated over the worldsheet to give BRST-invariant amplitudes.  Not
surprisingly, therefore, the new amplitudes obtained in [23] by including
screening charges are in on-to-one correspondence with amplitudes that we
can compute by including discrete states.  In particular, we have checked
that the scattering amplitude
$\fipf{\ft{15}{16}}{\ft{15}{16}}{\ft{15}{16}}{D}{\ft{15}{16}}$ for four
$\leff=\ft{15}{16}$ states (one from ghost-boosting (6.7) and three from
(6.8), as in [23]) and the discrete state $D=D^0[\ft27Q]$ agrees with the
result for the four-point  function
$\fipf{\ft{15}{16}}{\ft{15}{16}}{\ft{15}{16}}{\ft{15}{16}}{S}$ obtained  in
[23].

     In all the examples of scattering amplitudes involving discrete
states that we discussed in section 4, the discrete states could normal
order with external physical states to give new physical states. The new
result of [23] amounts to the observation that there are further
amplitudes that can be obtained, using the formalism that we presented in
[11] and this paper, by including discrete states that cannot normal order
with the external physical states. By this means, one can indeed  obtain
scattering amplitudes involving more than two   $\leff=\ft{15}{16}$
external states.  Our statements on this point in section 4 apply only to
the case where no discrete states are involved.

     Our formalism for  describing $W_3$ string scattering does not
suffer from any of the  deficiencies mentioned in [23], and indeed, since
the authors of [23] perform  all their calculations by using the
procedures that we have developed, (or, in the case of amplitudes
involving screening charges, a procedure that is  directly equivalent to
ours), their results necessarily agree with ours.

     We are grateful to the authors of [23] for encouraging us to look
further  at scattering amplitudes involving discrete states.  S.S. is
grateful to Jose M. Figueroa-O'Farrill for discussions about the screening
charge.

\np
\noindent
\centerline{\bf APPENDIX}
\bigskip
In this appendix we present the explicit forms of the various prime physical
operators that we use in this paper.  In cases where there is a freedom to
add BRST-trivial pieces, this is always precisely sufficient to enable us to
write the operators so that they admit an effective spacetime
interpretation.  It is in this form that we present the operators.  We first
list the physical operators with continuous momentum $p_\mu$, and then those
with discrete momentum.

\bigskip
\noindent{\bf Prime operators with continuous momentum}
\medskip

\noindent$\bullet\ \underline{\ell=0,\ G=3}$:
$$
\eqalignno{
\V{3}{1}{-\ft87 Q}{p} &=c\, \del \gamma\, \gamma\, e^{-\ft87 Q \varphi}
e^{i p\cdot X}\ ,&(A.1)\cr
\V{3}{1}{-\ft67 Q}{p} &=c\, \del \gamma\,\gamma \, e^{-\ft67 Q \varphi}
e^{i p\cdot X}\ ,
&(A.2)\cr
\V{3}{15/16}{-Q}{p} &=c\, \del\gamma\,\gamma \, e^{-Q \varphi}
e^{i p\cdot X}\ .&(A.3) \cr}
$$

\noindent$\bullet\ \underline{\ell=1,\ G=2}$:
$$
\eqalignno{
\V{2}{15/16}{-\ft37 Q}{p} &=\Big(c\,\gamma + {i \over 3
\sqrt{58}}\del\gamma\,  \gamma\Big)\, e^{-\ft37 Q \varphi} e^{ip\cdot X}\ ,
&(A.4)\cr \V{2}{1/2}{-\ft47 Q}{p} &=\Big(c\,\gamma - {i \over 3
\sqrt{58}}\del\gamma\,  \gamma\Big)\, e^{-\ft47 Q \varphi} e^{ip\cdot X}
\ .&(A.5)\cr}
$$

\noindent$\bullet\ \underline{\ell=2,\ G=2}$:
$$
\eqalign{
\V{2}{1/2}{-\ft27 Q}{p} = &{-i \over \sqrt{29}}
\Big(\del\varphi\,\del\gamma\,\gamma
+ \sqrt{58}\, i\,\del\varphi \,c\,\gamma
-\ft32 \sqrt{29}\, i\, c\,\del\gamma \cr
&-\ft23 \sqrt{2}\,
\del^2\gamma\,\gamma - \ft13 \sqrt{2}\,\, b\,c\,\del\gamma\,\gamma\Big)
e^{-\ft27 Q \varphi} e^{i p\cdot X}\ .\cr}\eqno(A.6)
$$

\noindent$\bullet\ \underline{\ell=3,\ G=1}$:
$$
\V{1}{1}{0}{p}=
\Big(c -\ft8{261}b\, \del\gamma\, \gamma -\ft4{87} \sqrt{29}\, i \,
\del\varphi\, \gamma +\ft7{174}\sqrt{58}\, i\, \del\gamma
\Big)e^{ip\cdot X}\ .
\eqno(A.7)
$$

\noindent$\bullet\ \underline{\ell=4,\ G=1}$:
$$
\eqalignno{
\V{1}{15/16}{\ft17Q}{p} = \Big(& b\, c\, \del\gamma + \ft5{348} \sqrt{58}\, i
\, b\, \del^2\gamma\, \gamma -\ft{27}{16}\beta\, \del\gamma\, \gamma
-\ft9{16}\sqrt{58}\, i\, c\, \beta\, \gamma -\ft12\sqrt{2}\, \del\varphi \,
b\, c\, \gamma\cr
& +\ft1{58}\sqrt{29}\, i\, \del\varphi\, b\, \del\gamma\, \gamma -\ft34
\sqrt{29}\, i\, \del\varphi\, c - \del\varphi\, \del\varphi\, \gamma
+\ft98\sqrt{2}\, \del\varphi\, \del\gamma +\ft1{16} \del b\, c\, \gamma
\cr
&+\ft{35}{928}\sqrt{58}\, i\, \del b\, \del\gamma\, \gamma -\ft34\sqrt{2}\,
\del^2\varphi\, \gamma +\ft12 \del^2\gamma \Big)
e^{\ft17 Q\varphi} e^{ip\cdot X}\ .&(A.8)\cr}
$$

\noindent$\bullet\ \underline{\ell=5,\ G=1}$:
$$
\eqalignno{
\V{1}{1}{\ft27Q}{p} = \Big( & b\, c\, \beta\, \del\gamma\, \gamma -\ft2{15}
b\, c\, \del^2\gamma -\ft2{1305} \sqrt{58}\, i\, b\,\del^2\gamma\,\del\gamma
-\ft2{783} \sqrt{58} \, i\, b\,\del^3\gamma\,\gamma
+\ft25\beta\,\del^2\gamma\,\gamma\cr
&-\ft3{20} \sqrt{58}\, i\, c\,\beta\,\del\gamma + \ft3{10}\sqrt{58}\, i\,
c\,\del\beta\,\gamma -\ft{11}{15}\sqrt{2}\, \del\varphi\, b\, c\,\del\gamma -
\ft4{435}\sqrt{29}\, i\,\del\varphi\, b\,\del^2\gamma\,\gamma \cr
&+\ft9{10} \sqrt{2}\,\del\varphi\,\beta\,\del\gamma\,\gamma +\ft35 \sqrt{29}
\, i\,\del\varphi\, c\, \beta\,\gamma+\ft8{15} \del\varphi\,\del\varphi\,
b\, c\,\gamma-\ft8{1305}\sqrt{58}\, i\, \del\varphi\,\del\varphi\,
b\,\del\gamma\,\gamma \cr
& + \ft15\sqrt{58}\, i\, \del\varphi\,\del\varphi\, c + \ft4{15}\sqrt{2}\,
\del\varphi\,\del\varphi\,\del\varphi\,\gamma -\ft{11}{15}
\del\varphi\,\del\varphi\, \del\gamma +\ft16 \sqrt{2}\, \del\varphi\,\del b \,
c\,\gamma \cr
&-\ft{91}{2610}\sqrt{29}\, i\, \del\varphi\, \del b\,\del\gamma\, \gamma
-\ft4{15} \sqrt{2}\, \del\varphi\,\del^2\gamma -\ft{13}{870} \sqrt{58}\, i\,
\del b\, b\, c\,\del\gamma\,\gamma -\ft5{12}\del b\, c\,\del\gamma\cr
&-\ft{19}{1305} \sqrt{58}\, i\, \del b\, \del^2\gamma\, \gamma +\ft{11}{10}
\del\beta\,\del\gamma\,\gamma +\ft13 \sqrt{2}\, \del^2 \varphi\, b\, c\,\gamma
- \ft1{145}\sqrt{29}\, i\, \del^2\varphi\, b\,\del\gamma\,\gamma \cr
&+\ft1{10}\sqrt{29}\, i\,\del^2\varphi\, c +\ft{14}{15} \del^2\varphi\,
\del\varphi\,\gamma -\ft{23}{60}\sqrt{2}\, \del^2\varphi\,\del\gamma -\ft1{30}
\del^2b\, c\,\gamma\cr
&-\ft{19}{1044} \sqrt{58}\, i\, \del^2 b\,\del\gamma\,\gamma +\ft15 \sqrt{2}\,
\del^3\varphi\,\gamma -\ft2{45} \del^3\gamma\Big) e^{\ft27 Q\varphi}
e^{ip\cdot X}\ . &(A.9) \cr}
$$

\noindent$\bullet\ \underline{\ell=8,\ G=0}$:
$$
\eqalignno{
&\V{0}{1/2}{\ft47 Q}{p}=\cr
&\Big(
\ft{20}{261}\sqrt{2}\,  b\, \beta\, \del^3\gamma\, \gamma
-\ft{10}{29}\sqrt{2}\,  b\, \beta\, \del^2\gamma\,   \del\gamma+
\ft{4}{29}\sqrt{29} \,i\,  b\,  c\, \beta\, \del^2\gamma-
\ft4{87}\sqrt{29} \,i\,  b\,  c\, \del\beta\,   \del\gamma\cr
&+ \ft2{29}\sqrt{29} \,i\,  b\,  c\, \del^2\beta\, \gamma- \ft{20}{261}
\sqrt{2}\,  b\, \del\beta\,  \del^2\gamma\, \gamma + \ft{17}{87}
\sqrt{2}\,  b\, \del^2\beta\,  \del\gamma\, \gamma+
\ft1{522}\sqrt{2}\, b\, \del^4\gamma\cr
&+ \ft{4}{87}\sqrt{29}\,i\, \beta\, \del^3\gamma+   3 \sqrt{2}\,  c\,
\del\beta\, \beta\, \gamma- \ft34\sqrt{2}\,  c\, \del^2\beta+
\ft{16}{87} \del\varphi\,  b\, \beta\, \del^2\gamma\, \gamma+
\ft{4}{87}\sqrt{29}\,i\, \del\varphi\,  b\,  c\, \del\beta\, \gamma
\cr
& + \ft{32}{87} \del\varphi\,  b\, \del\beta\,   \del\gamma\,
\gamma- \ft{52}{783} \del\varphi\,  b\, \del^3\gamma+ \del\varphi\,  c\,
\del\beta  - \ft{16}{261} \del\varphi\, \del\varphi\, \del\varphi\,\del
b\,\gamma - \ft{32}{261}\sqrt{2}\,\del\varphi\, \del\varphi\, \del b\,
b\,  c\, \gamma\cr
&- \ft{32}{22707} \sqrt{29}\,i\, \del\varphi\, \del\varphi\, \del b\,
b\, \del\gamma\, \gamma - \ft4{87}\sqrt{29}\,i\,\del\varphi\,
\del\varphi\, \del b\,  c+ \ft{10}{87}\sqrt{2}\,\del\varphi\,
\del\varphi\, \del b\,   \del\gamma \cr
&+ \ft4{87}\sqrt{29}\,i\,\del\varphi\, \del\varphi\, \del\beta\, \gamma +
\ft{2}{87}\sqrt{2}\, \del\varphi\, \del\varphi\,  \del^2 b\, \gamma+
\ft{104}{261}\del\varphi\, \del b\, b\,  c\, \del\gamma+
\ft{64}{7569}\sqrt{58}\,i\,\del\varphi\, \del b\, b\, \del^2\gamma\,
\gamma\cr
&- \ft{28}{87} \del\varphi\, \del b\,  \beta\, \del\gamma\, \gamma-
\ft4{29}\sqrt{58}\,i\, \del\varphi\, \del b\,  c\, \beta\, \gamma   -
\ft5{58}\sqrt{58}\,i\, \del\varphi\, \del\beta\, \del\gamma - \ft8{261}
\del\varphi\, \del^2 b\, b\,  c\, \gamma\cr
&+  \ft{32}{7569}\sqrt{58}\,i\, \del\varphi\, \del^2 b\, b\,
\del\gamma\, \gamma+ \ft1{58}\sqrt{58}\,i\, \del\varphi\, \del^2 b\,  c-
\ft{23}{174} \del\varphi\, \del^2 b\, \del\gamma- \ft1{29}\sqrt{58}\,i\,
\del\varphi\, \del^2\beta\, \gamma\cr
& - \ft1{87}\del\varphi\, \del^3 b\, \gamma- \ft8{29}\sqrt{2}\, \del b\,
b\,  c\, \beta\, \del\gamma\, \gamma + \ft4{87}\sqrt{2}\, \del b\,
b\,  c\, \del^2\gamma- \ft{172}{7569}\sqrt{29}\,i\, \del b\,
b\, \del^2\gamma\, \del\gamma\cr
&+ \ft{268}{68121} \sqrt{29}\,i\, \del b\,  b\, \del^3\gamma\, \gamma+
\ft8{87}\sqrt{2}\, \del b\, \beta\, \del^2\gamma\, \gamma+
\ft8{29}\sqrt{29} \,i\, \del b\,  c\,  \beta\, \del\gamma-
\ft7{87}\sqrt{29}\,i\,\del b\, c\, \del\beta\, \gamma\cr
&- \ft{67}{174}\sqrt{2}\, \del b\, \del\beta\,  \del\gamma\, \gamma
+ \ft2{87}\sqrt{2}\, \del b\, \del^3\gamma+ \ft{15}{29}\sqrt{29}\,i\,
\del\beta\, \beta\,  \del\gamma\, \gamma- \ft{2}{87}\sqrt{29}\,i\,
\del\beta\, \del^2\gamma\cr
&+ \ft8{87} \del^2\varphi\,  b\, \beta\, \del\gamma\, \gamma+
\ft{68}{261} \del^2\varphi\,  b\, \del^2\gamma+ \ft{15}{58}\sqrt{58} \,i\,
\del^2\varphi\, \beta\, \del\gamma  - 3 \del^2\varphi\,  c\, \beta\,   +
\ft8{87}\sqrt{29} \,i\, \del^2\varphi\, \del\varphi\,  b\,  c\cr
&-  \ft{20}{87}\sqrt{2}\, \del^2\varphi\, \del\varphi\,  b\,
\del\gamma- \ft4{29}\sqrt{29}\,i\, \del^2\varphi\, \del\varphi\, \beta\,
\gamma+ \ft{32}{261} \del^2\varphi\, \del\varphi\, \del\varphi\,  b\,
\gamma - \ft{10}{87}\sqrt{2}\, \del^2\varphi\, \del\varphi\, \del
b\, \gamma  \cr
&-  \ft{40}{261} \del^2\varphi\, \del b\, b\,  c\, \gamma-
\ft{8}{7569}\sqrt{58}\,i\,\del^2\varphi\, \del b\,  b\, \del\gamma\,
\gamma+ \ft3{58}\sqrt{58} \,i\, \del^2\varphi\, \del b\,  c+  \ft{17}{58}
\del^2\varphi\, \del b\, \del\gamma\cr
&+ \ft{4}{29}\sqrt{58}\,i\,  \del^2\varphi\, \del\beta\, \gamma   +
\ft4{29}\sqrt{29} \,i\, \del^2\varphi\, \del^2\varphi+ \ft8{87}
\del^2\varphi\, \del^2 b\, \gamma+ \ft2{29}\sqrt{2}\, \del^2 b\,
b\,  c\, \del\gamma\cr
&+ \ft{20}{22707}\sqrt{29}\,i\,\del^2 b\,  b\, \del^2\gamma\, \gamma+
 \ft{13}{58}\sqrt{2}\, \del^2 b\, \beta\,  \del\gamma\, \gamma+
\ft1{29}\sqrt{29}\,i\, \del^2 b\,  c\, \beta\, \gamma-
\ft1{174}\sqrt{2}\,\del^2 b\, \del b\,  c\, \gamma\cr
&- \ft{167}{15138} \sqrt{29} \,i\, \del^2 b\, \del b\, \del\gamma\,
\gamma+  \ft4{87}\sqrt{2}\, \del^2 b\, \del^2\gamma +
\ft{15}{116}\sqrt{29} \,i\, \del^2\beta\, \del\gamma+
\ft{1}{29}\sqrt{58}\,i\,  \del^3\varphi\,  b\,  c\,  \cr
& - \ft{53}{261} \del^3\varphi\,  b\, \del\gamma - \ft2{29}\sqrt{58}\,i
\del^3\varphi\, \beta\, \gamma + \ft{28}{261}\sqrt{2}\, \del^3\varphi\,
\del\varphi\,  b\, \gamma- \ft2{87}\sqrt{29}\,i\, \del^3\varphi\,
\del\varphi \cr
&- \ft{14}{261} \del^3\varphi\, \del b\, \gamma+ \ft7{783} \sqrt{2}\,
\del^3 b\,  b\,  c\, \gamma
+ \ft{331}{68121}\sqrt{29} \,i\, \del^3 b\,  b\, \del\gamma\,
\gamma- \ft1{1044}\sqrt{29}\,i\, \del^3 b\,  c \cr
&+ \ft{23}{696}\sqrt{2}\, \del^3 b\, \del\gamma+ \ft8{261}
\del^4\varphi\,  b\, \gamma+ \ft1{87}\sqrt{58}\,i\, \del^4\varphi\, +
\ft1{522}\sqrt{2}\,\del^4 b\, \gamma\Big)
e^{\ft47 Q}e^{i p\cdot X} \ .&(A.10)\cr}
$$

\bigskip
\noindent{\bf Prime operators with discrete momentum}
\medskip

\noindent$\bullet\ \underline{\ell=1,\ G=2}$:
$$
\eqalignno{
\D{2}{-\ft67Q}&= \Big(c\, \gamma -{i\over 3\sqrt{58}} \del\gamma\, \gamma
\Big) e^{-\ft67 Q\varphi}\ ,&(A.11)\cr
\D{2}{-\ft87Q}&= \Big(c\, \gamma +{i\over 3\sqrt{58}} \del\gamma\, \gamma
\Big) e^{-\ft87 Q\varphi}\ .&(A.12)\cr}
$$

\noindent$\bullet\ \underline{\ell=4,\ G=0}$:
$$
\D00=1\ .\eqno(A.13)
$$

\noindent$\bullet\ \underline{\ell=6,\ G=0}$:
$$
\eqalign{
\D0{\ft27Q}&=
\Big(b\, \beta\, \del\gamma\, \gamma -\ft35\sqrt{58}\,i\, b\, c\, \beta\,
\gamma-\ft{14}{15}b\, \del^2\gamma - \ft{21}{20}\sqrt{58}\,i\,
\beta\,\del\gamma
+\ft{261}{10}c\, \beta\cr
&-\ft65 \sqrt{29}\,i\, \del\varphi\, b\, c - \ft{11}{15}\sqrt{2}\,
\del\varphi\, b\, \del\gamma -\ft65\sqrt{29}\,i\, \del\varphi\, \beta\,
\gamma -\ft45\sqrt{58}\,i\, \del\varphi\, \del\varphi +\ft8{15}
\del\varphi\, \del\varphi\, b\, \gamma\cr
&+\ft1{15}\sqrt{2}\, \del\varphi\, \del b\, \gamma -\ft15\del b\, b\, c\,
\gamma +\ft7{290}\sqrt{58}\,i\, \del b\, b\, \del\gamma\, \gamma -
\ft9{20}\sqrt{58}\,i\, \del b\, c -\ft{97}{60}\del b\, \del\gamma\cr
&-\ft65 \sqrt{58}\,i\, \del\beta\, \gamma -\ft85 \sqrt{29}\,i\, \del^2
\varphi + \ft8{15}\sqrt{2}\, \del^2\varphi\, b\, \gamma -\ft8{15} \del^2 b\,
\gamma \Big)e^{\ft27 Q\varphi}\ .\cr}\eqno(A.14)
$$

\np
\singlespace
\centerline{\bf REFERENCES}
\frenchspacing
\bigskip

\item{[1]}A.B. Zamolodchikov, {\sl Teor. Mat. Fiz.} {\bf 65} (1985)
1205.

\item{[2]}A. Bilal and J.-L. Gervais, {\sl Nucl. Phys.} {\bf B314} (1989)
646.

\item{[3]}S.R. Das, A. Dhar and S.K. Rama, {\sl Mod. Phys. Lett.}
{\bf A6} (1991) 3055; {\sl Int. J. Mod. Phys.} {\bf A7} (1992) 2295.

\item{[4]}C.N. Pope, L.J. Romans and K.S. Stelle, {\sl Phys.
Lett.} {\bf B268} (1991) 167; {\sl Phys. Lett.} {\bf B269} (1991) 287.

\item{[5]}C.N. Pope, L.J. Romans, E. Sezgin and K.S. Stelle,
{\sl Phys. Lett.} {\bf B274} (1992) 298.

\item{[6]}H. Lu, C.N. Pope, S. Schrans and K.W. Xu,  {\sl Nucl.
Phys.} {\bf B385} (1992) 99.

\item{[7]}H. Lu, C.N. Pope, S. Schrans and X.J. Wang,  {\sl Nucl.
Phys.} {\bf B379} (1992) 47.

\item{[8]}H. Lu, B.E.W. Nilsson, C.N. Pope, K.S. Stelle and P.C. West,
``The low-level spectrum of the $W_3$ string,'' preprint CTP TAMU-64/92,
hep-th/9212017.

\item{[9]}C.N. Pope, E. Sezgin, K.S. Stelle and X.J. Wang, ``Discrete
states in the $W_3$ string,'' preprint, CTP TAMU-64/92, hep-th/9209111, to
appear in {\sl Phys. Lett}. {\bf B}.

\item{[10]}P.C. West, ``On the spectrum, no ghost theorem and modular
invariance of $W_3$ strings,'' preprint KCL-TH-92-7, hep-th/9212016.

\item{[11]}H. Lu, C.N. Pope, S. Schrans and X.J. Wang,  ``The interacting
$W_3$ string,'' preprint CTP TAMU-86/92, KUL-TF-92/43, hep-th/9212117.

\item{[12]}C.N. Pope and X.J. Wang, ``The ground ring of the $W_3$ string,''
preprint CTP-TAMU-3/93, to appear in the proceedings of the Summer School in
High-Energy Physics and Cosmology, Trieste 1992.

\item{[13]}S.K. Rama, {\sl Mod.\ Phys.\ Lett.}\ {\bf A6} (1991) 3531.

\item{[14]}B.H. Lian and G.J. Zuckermann, {\sl Phys. Lett.}\ {\bf B254} (1991)
 541.

\item{[15]}E. Witten, {\sl Nucl. Phys.} {\bf B373} (1992) 187;\nl
E. Witten and B. Zwiebach, {\sl Nucl. Phys.} {\bf B377} (1992) 55.

\item{[16]}J. Thierry-Mieg, {\sl Phys. Lett.} {\bf B197} (1987) 368.

\item{[17]}L.J.  Romans, {\sl Nucl.  Phys.} {\bf B352} (1991) 829.

\item{[18]}K.S. Stelle, ``$W_3$ strings,'' in proceedings of the
Lepton-Photon/HEP Conference, Gen\`eve 1991 (World Scientific, Singapore,
1992).

\item{[19]}M.E. Peskin, ``Introduction to string and superstring theory
II,'' lectures at the 1986 Theoretical Advanced Study Institute in Particle
Physics, Santa Cruz.

\item{[20]}A.A. Belavin, A.M. Polyakov and A.B. Zamolodchikov, {\sl Nucl.
Phys.} {\bf B241} (1984) 333.

\item{[21]}M.D. Freeman and P.C. West, ``$W_3$ string scattering,''
preprint, KCL-TH-92-4, hep-th/9210134.

\item{[22]}K. Thielemans, {\sl Int. J. Mod. Phys.} {\bf C2} (1991) 787.

\item{[23]}M.D. Freeman and P.C. West, ``The covariant scattering and
cohomology of $W_3$ strings,'' preprint, KCL-TH-93-2, hep-th/9302114.

\end